\documentclass[
amsmath, 
amssymb, 
aps, 
pra,
twocolumn,
floatfix,
nofootinbib
]{revtex4}
\usepackage[utf8]{inputenc}
\usepackage{titlesec}
\usepackage[colorlinks]{hyperref}
\usepackage{amsmath,amssymb}
\usepackage{bbold}
\usepackage{float}
\usepackage{dcolumn}% Align table columns on decimal point
\usepackage{bm}% bold math
\usepackage{dsfont}
\usepackage{url}
\usepackage{mathtools}

\usepackage{amsfonts}
\usepackage{braket}
\usepackage{cancel}
\usepackage{bbold}
\usepackage{makecell}
\usepackage[normalem]{ulem}
\usepackage{xcolor}
\usepackage{comment}
\newcommand{\tr}{\text{tr}}
\usepackage{afterpage}

\begin{document}
\title{Forty Thousand Kilometers Under Quantum Protection}

\author{N.\,S.\,Kirsanov$^1$}

\author{V.\,A.\,Pastushenko$^{1,\dag}$}

\author{A.\,D.\,Kodukhov$^{1,\dag}$}

\author{M.\,V.\,Yarovikov$^1$}

\author{A.\,B.\,Sagingalieva$^1$}

\author{D.\,A.\,Kronberg$^1$}

\author{M.\,Pflitsch$^1$}

\author{V.\,M.\,Vinokur$^{1,*}$}

\address{$^1\,$Terra Quantum AG, Kornhausstrasse 25, 9000 St.\,Gallen, Switzerland
\\$^\dag\,$these authors contributed equally to this work
\\$*\,$vv@terraquantum.swiss
}

\begin{abstract}
Quantum key distribution (QKD) is a revolutionary cryptography response to the rapidly growing cyberattacks threat posed by quantum computing.
Yet, the roadblock limiting the vast expanse of secure quantum communication is the exponential decay of the transmitted quantum signal with the distance.
Today's quantum cryptography is trying to solve this problem by focusing on quantum repeaters.
However, efficient and secure quantum repetition at sufficient distances is still far beyond modern technology.
Here, we shift the paradigm and build the long-distance security of the QKD upon the quantum foundations of the Second Law of Thermodynamics and end-to-end physical oversight over the transmitted optical quantum states.
Our approach enables us to realize quantum states' repetition by optical amplifiers keeping states' wave properties and phase coherence.
The unprecedented secure distance range attainable through our approach opens the door for the development of scalable quantum-resistant communication networks of the future.
\end{abstract}
     
\maketitle

\section*{Introduction}
The quantum threat to secure communications makes top headlines and Niagara Falls of reviews and research explaining how quantum computers using, for example, Shor’s algorithm\,\cite{Shor}, devalue the existing cryptographic schemes. 
Remarkably, the same advances in quantum physics that have created this quantum threat enable solutions for quantum security.
Building upon quantum phenomena, novel quantum cryptography offers methods for unparalleled security, including quantum secure direct communication\,\cite{QSDC1,QSDC2,QSDC3,QSDC4,QSDC5,QSDC6}, probabilistic one-time programs\,\cite{One_time2013, One_time2018, roehsner2021probabilistic}, and quantum key distribution (QKD)\,\cite{BB84, Ekert, B-92, Six-State, Advances_QKD,MDI2022,MDI2022_400km}.
The QKD, on which we focus in our present work, allows two parties to share a secret bit sequence for various applications.  
The existing QKD protocols are efficient over relatively short distances due to the fundamental Pirandola-Laurenza-Ottaviani-Banchi (PLOB) bound\,\cite{PLOB}, which dictates that secret communication rates decrease exponentially with channel length.
The simplest approach to resolve this issue is to use the trusted reproduction nodes along the transmission line\,\cite{TrustedNodesSatellite, TrustedNodes4600, TrustedNodesNature}, which is a compromise to the overall security.
The alternative solution is the utilization of quantum repeaters\,\cite{Kimble_Qrepeater, Sangouard_Qrepeater, Simon_Qrepeater, Duan_Qrepeater, Briegel_Qrepeater, Kok_Qrepeater, Childress_Qrepeater, van_Loock_Qrepeater, Wang_Qrepeater, Sangouard2_Qrepeater, Azuma_Qrepeater, Zwerger_Qrepeater, Munro_Qrepeater, Jiang_Qrepeater, Munro2_Qrepeater, Grudka_Qrepeater} which eliminates the need for trust in the intermediate relay.
However, since quantum repetition manipulates fragile entangled states, its implementation at a long scale remains beyond state-of-the-art technologies.

Here, contemplating the physical nature of the quantum states' transmission, we lift the PLOB bound by using restrictions of quantum thermodynamics and the end-to-end physical control over losses in the optical quantum channel.
We shift the quantum cryptography paradigm building on the same quantum considerations that provide the foundations of the Second Law of Thermodynamics.
Our approach ensures signal repetition through optical amplification, presumes no trust at the intermediate channel points, and expands the secure transmission range to global distances.
This paves the way for constructing scalable quantum communication networks of the future---a problem that has garnered significant interest in recent years\,\cite{Network2021,yin2023experimental,TF2023}.

\section*{General idea}
Conventionally, the eavesdropper (Eve) is seen as capable of exploiting all the losses from the transmission channel, irrespective of their origin.
This puts a strong restriction on the number of photons in the transmitted quantum states, which significantly complicates their repetition.
However, upon close quantum mechanical examination, this presupposition appears unrealistic.
In reality, the majority of losses in optical fibers occur due to the light scattering on the quenched disorder and are distributed homogeneously along the line (hereinafter, we will be referring to such losses as to natural losses).
In a single mode silica fiber's 1530--1565\,nm wavelength window, the standard for modern telecommunications, these losses amount to approximately $4\times 10^{-5}$ of the passing signal's intensity per meter. 

We describe the information dynamics of the randomized signal transmitted over an optical channel. 
This consideration is carried out analogously to consideration of the Second Law of Thermodynamics, i.e., the dynamics of entropy, through the lens of the microscopic quantum mechanical laws\,\cite{Lesovik_Qirreversibility, Kirsanov_Qirreversibility, Lesovik2_Qirreversibility}.
Had the system been isolated, its entropy would not decrease, i.e., Eve would not be able to obtain any information.
In the presence of natural losses, the system can no longer be regarded as isolated, and thus, the eavesdropper gets an opportunity to decrease the system's entropy in analogy with the quantum Maxwell demon.
However, in order to glean information from the scattering losses of relatively weak signals that we employ for our approach, Eve has to use quantum detection devices spanning an unfeasible length of optical fiber, see Supplementary Note\,1.
That is why one concludes that in this weak signal regime, Eve is unable to effectively collect and exploit natural losses.

Losses other than natural ones can, in turn, be physically controlled.
We propose a technique of physical loss control (line tomography) implying that legitimate users detect local interventions by comparing the constantly updated tomogram of the line with the initial one, knowingly obtained in the absence of Eve.
Line tomography involves sending the high-frequency test light pulses and analyzing their reflected (via the technique known as the time-domain reflectometry\,\cite{OTDR}) and the transmitted components.
The coupling of photons to any eavesdropping system is impossible without modifying the fiber medium, which in turn inevitably changes the line tomogram.
Unable to perform such radical interventions unnoticed, Eve is thus restricted to introducing small local leakages, which are precisely measured by the users.
This implicates the possibility of employing the information-carrying light states containing the numbers of photons that are sufficient to repeat the states through optical amplification yet not enough to be easily eavesdropped on. 

Utilizing a cascade of accessible optical amplifiers to counteract the degradation of signals over extensive distances, as opposed to the employment of quantum repeaters\,\cite{Kimble_Qrepeater, Sangouard_Qrepeater, Simon_Qrepeater, Duan_Qrepeater, Briegel_Qrepeater, Kok_Qrepeater, Childress_Qrepeater, van_Loock_Qrepeater, Wang_Qrepeater, Sangouard2_Qrepeater, Azuma_Qrepeater, Zwerger_Qrepeater, Munro_Qrepeater, Jiang_Qrepeater, Munro2_Qrepeater, Grudka_Qrepeater}, enables global transmission and high key distribution rates.
It is important to note that these optical amplifiers should not be viewed as trusted nodes, as the integrity of the transmission scheme is maintained through end-to-end control by legitimate users, and there is no recourse to the form of classical data.

We showcase our approach via a prepare-and-measure QKD protocol utilizing non-orthogonal coherent photonic states $\ket{\gamma_0}$ and $\ket{\gamma_1}$ for encoding 0 and 1 bits.
In the protocol's framework, our approach means restricting the fraction of photons leaked to Eve, $r_\text{E}$, to ensure that the leaked states $\ket{\sqrt{r_\text{E}}\gamma_{0}}$ and $\ket{\sqrt{r_\text{E}}\gamma_{1}}$ sufficiently overlap, i.e., $\left|\left\langle \sqrt{r_\text{E}}\gamma_{0}\big|\sqrt{r_\text{E}}\gamma_{1}\right\rangle\right|\sim 1$ (these states become mixed if the transmission channel includes amplifiers, but for now we ignore this fact for the sake of simplicity). 
Eve cannot by any means---except by completely blocking part of the signal pulses\,\cite{USD} but this is prevented by the line tomography---extract more information than the Holevo quantity $\chi_\text{E}$\,\cite{Holevo} which tends to zero when $\left|\left\langle \sqrt{r_\text{E}}\gamma_{0}\big|\sqrt{r_\text{E}}\gamma_{1}\right\rangle\right|\to 1$.
The users monitor the value of $r_\text{E}$ and adapt the parameters of $\ket{\gamma_0}$ and $\ket{\gamma_1}$ to ensure that the intercepted pulses are poorly distinguishable.

Thus, we overcome the PLOB bound by what can be called the ``channel device-dependent'' approach. 
This approach is no less physically justified as a traditional device-dependent scenario where the eavesdropper is assumed not to be able to substitute some of the equipment at the sender and receiver side. 
Hence, there are no compromises in security that allow us to increase the secret key distribution distance, only higher device dependence, with correct channel work ensured by tomography methods.

\section*{Protocol description}
We put forth an exemplary QKD protocol based on our physical control approach.
Let the legitimate users, the sender, Alice, and the receiver, Bob, be connected via a classical authenticated communication channel and optical line serving as a quantum channel.
The protocol is designed as follows:
\begin{enumerate}
    \item[]\textbf{Initial preparation}
    \item [0.] Alice and Bob carry out initial line tomography to determine the natural losses that Eve cannot exploit.
    At this and only this preliminary step, the legitimate users must be certain that Eve has no influence on the line.
    The users share the tomogram via the classical channel. 

    \item[] \textbf{Line tomography}
    \item Alice and Bob perform the physical loss control over the line and, through comparison with the initial line tomogram, infer the fraction $r_\text{E}$ of the signal possibly seized and exploited by Eve.
    The users also localize the points of Eve's intervention.
    To update the line tomogram, users exchange information via the classical channel.
    If the stolen fraction grows too large so that the evaluated legitimate users' information advantage over Eve disappears---the analytical estimate for this advantage is provided below---the transmission is terminated.

    \item[] \textbf{Transmission of quantum states}
    
    \item Using a random number generator, Alice produces a bit sequence of the length $L$.
    Alice ciphers her bit sequence into a series of $L$ coherent light pulses, which she sends to Bob.
    The bits 0 and 1 are encoded into coherent states $\ket{\gamma_0}$ and $\ket{\gamma_1}$ respectively. 
    Their parameters are optimized based on the known fraction of the signal seized by Eve $r_\text{E}$ and Eve's position in the line. 
    The optimal parameters correspond to the maximum key distribution rate at given losses in the channel, the analytical relation for which is presented below.
    The optimal parameters are considered to be known to Alice and Bob and also to Eve.
    
    \item The signals are amplified by the cascade of optical amplifiers installed along the optical line, possibly equidistantly.
    Bob receives the signals and measures them.

    \item[] \textbf{Classical post-procession}
    
    \item Alice and Bob perform the postselection, i.e., they discard the positions corresponding to corrupted measurement outcomes. 
    The postselection criteria are defined by the set of parameters, which are optimally calculated by the users.
    
    \item The users perform error correction.
    The procedure can be done with well-known classical methods, e.g., linear codes\,\cite{MacKay, LDPC2, Hamming}, or with methods designed specifically for the QKD, such as the Cascade protocol\,\cite{Cascade1, Cascade2, Cascade3}.
    
    \item The users estimate Eve's information obtained at the previous stages and perform the privacy amplification procedure to produce a shorter key (e.g., using the universal hashing method\,\cite{Carter_universal_hashing}) on which Eve has none or negligibly small information.
\end{enumerate}
The steps from 1 to 6 directly constitute the process of key distribution, and the users must repeat them until satisfied with the total shared key length.
It is important to note that the key is not generated solely by Alice in step 2; instead, it emerges through the collaborative simultaneous actions of both parties, with postprocessing playing a vital role in the process.

The particular way of encoding bits 0 and 1 into the parameters of coherent pulses $\ket{\gamma_0}$ and $\ket{\gamma_1}$ may vary.
For illustrative purposes, we concentrate on the two simplest and straightforward schemes, viz. encoding bits into pulses with (a) different photon numbers, $|\gamma_0|^2 \neq|\gamma_1|^2$, and phase randomization\,\cite{PhR_molmer1997,PhR_lutkenhaus, PhR_van}, and (b) same photon numbers, $|\gamma_0|^2 =|\gamma_1|^2$, and phases different by $\pi$.
In both encoding schemes, Alice varies $|\gamma_0|^2$ and $|\gamma_1|^2$. 
More sophisticated encoding schemes, for instance, schemes leveraging the pulses' shapes, make exploiting the natural scattering losses even more unsolvable.
To complicate the problem further, the cable design may include an encapsulating layer of metal of heavily doped silica, transforming the scattering radiation into heat under the control of the users; see Supplementary Note 2 for details.

\section*{Protocol security}
Here, we delve into the security of the described protocol, by building upon the following:
\begin{enumerate}
    \item Alice and Bob each generate random numbers that Eve cannot predict.
    \item Other from the transmission channel---which is a fiber line with the embedded optical amplifiers---and the classical authenticated channel, users' equipment is isolated from Eve.
    \item Eve cannot effectively collect and exploit natural losses from the transmission channel.
    To eavesdrop on the signal, Eve must introduce new artificial local leakages. 
    Eve can also use the local leakages on the original fiber discontinuities, such as bends or connections. 
    \item The transmission line between Alice and Bob is characterized by the initial line tomogram. 
    All losses constituting deviations from the initial tomogram are attributed to Eve.
    \item Eve is bound to the beam-splitting attack.
    She may seize some fraction of the signal at any point of the optical line. 
    With that, she is unable to replace any section of the line with a channel of her own creation since this action is detectable by the line tomography.
\end{enumerate}

Attacks that deviate from the beam-splitting attack necessitate a significant alteration of the line tomogram, in which case the protocol should be terminated; as such, we will not delve into them here. 
We will refer to the point of Eve's intrusion into the line as the ``beam splitter'' and assume that any reflection back towards Alice from this point is insignificant.
As our analysis will demonstrate, the protocol's efficiency is contingent on Eve's placement along the line. 
For the sake of simplicity, we will focus on the scenario in which Eve intercepts from a single point at the line.
Indeed, with some overhead, the case where Eve intercepts from multiple points---stealing $r_\text{E}^{(i)}$ from the $i$-th local leakage---can be reduced to a situation where Eve is at the single worst location for the users among all identified interception points, and she steals the effective overall leakage 
\begin{equation}
\label{total_leak}
r_\text{E}=\sum_i r_\text{E}^{(i)}\prod_{0\leq j<i}(1-r_\text{E}^{(j)})=1-\prod_{i}(1-r_\text{E}^{(i)}),
\end{equation}
where we defined $r_\text{E}^{(0)}=0$.
This effective overall leakage is detectable by transmittometry, see Sec.\,``Physical loss control and amplification'' for details.
The detailed analysis of the multi-point interception and constructive interference will be the subject of our forthcoming publication.

To evaluate the protocol's security, we describe the evolution of Alice's, Bob's, and Eve's quantum systems and quantify the information available to different parties.
We examine the case in which the beam splitter is placed immediately following one of the amplifiers, as this arrangement is most advantageous for Eve, but, with minimal adjustments, the same analysis can be applied to any arbitrary beam splitter's position.
We derive an analytical expression for the length of the final secure key $L_\text{f}$, which represents the users' informational advantage over Eve given the fixed value of $r_\text{E}$ and the distance between Alice and Eve $D_{\text{AE}}$. 
This expression depends on the encoding and postselection parameters and should be maximized by the users to determine the parameters' optimal values.
The condition $L_{\text{f}}/L>0$ for the chosen parameters ensures successful secret key generation\,\cite{Devetak_Winter}.

At the beginning of the protocol, Alice encodes the logical bits into the coherent states with the different complex amplitudes, $0 \rightarrow \ket{\gamma_0},\,1 \rightarrow \ket{\gamma_1}$.
In the photon number encoding scheme, the pulses are different in the average numbers of photons $|\gamma_0|^2$ and $|\gamma_1|^2$,
while the phase of each pulse is random.
The photon number measurement at Bob's end is formalized in terms of the projective operators:
\begin{subequations}
\begin{equation}
\begin{gathered}
    \hat{E}_0 = \sum \limits_{k=\mu-\theta_3}^{\mu-\theta_1} \ket{k} \bra{k},\quad\hat{E}_1 = \sum \limits_{k=\mu+\theta_2}^{ \mu+\theta_4} \ket{k} \bra{k},\\\hat{E}_\text{fail} = \hat{\mathds{1}} - \hat{E}_0 - \hat{E}_1,
    \label{Eint}
\end{gathered}
\end{equation}
where $\hat{E}_0$, $\hat{E}_1$ and $\hat{E}_\text{fail}$ correspond to 0, 1, and failed---meaning that this result should be later discarded---outcomes respectively, $\ket{k}$ is the Fock state of $k$ photons, $\hat{\mathds{1}}$ is the identity operator, $\mu=\left(\left|\gamma_0\right|^2+\left|\gamma_1\right|^2\right)/2$, and $\theta_{1-4}$ are the postselection parameters tuned by Bob depending on the proportion of the stolen signal.
The photon numbers between $\mu-\theta_1$ and $\mu+\theta_2$ are difficult to relate to 0 or 1, while numbers below $\mu-\theta_3$ and above $\mu+\theta_4$ are associated with the information corruption: as we show in Supplementary Note 4, optical amplification imposes correlations between pulses received by Bob and Eve, and extreme photon numbers at Bob's end also constitute very distinguishable signals for Eve.

For the phase encoding (note that using this scheme requires that the optical fiber is phase-preserving), the pulses are characterized by the same average photon number, $|\gamma_0|^2=|\gamma_1|^2$, but by different, although fixed, phases.
For instance, the relative phase can be $\pi$, $\gamma_0=-\gamma_1=\gamma\in\mathds{R}$, and then, to distinguish the pulses, Bob should perform homodyne measurement of the quadrature $\hat{q}$ corresponding to the real axis in the phase space:
\begin{equation}
\begin{gathered}
        \hat{E}_0 = \int\limits^{\theta_{\text{2}}^\prime}_{\theta_{\text{1}}^\prime} d q \ket{q}\bra{q},\quad\hat{E}_1 = \int\limits^{-\theta_{\text{1}}^\prime}_{-\theta_{\text{2}}^\prime} d q \ket{q}\bra{q},\\\hat{E}_\text{fail} = \hat{\mathds{1}} - \hat{E}_0 - \hat{E}_1,
        \label{Ephase}
\end{gathered}
\end{equation}
\end{subequations}
where $\ket{q}$ is the eigenstate of $\hat{q}$, and $\theta_{\text{1,2}}^\prime$ play the same role as $\theta_{1-4}$ in the photon number encoding case.
With this scheme, we deal only with two postselection parameters because probability distributions of measurement results for two pulses are symmetric with respect to $q=0$.

For both encoding schemes, the operational values of $|\gamma_0|$, $|\gamma_1|$ and $\theta_{1-4}$ ($\theta_{\text{1,2}}^\prime$) are determined via maximizing the analytical expression for the predicted length of the final secure key $L_\text{f}$, which, in turn, depends on the proportion of the stolen signal $r_\text{E}$ and the distance between Alice and Eve $D_\text{AE}$.
Correlations between the states at Eve's and Bob's disposal due to optical amplification drastically complicate the analytical description of states' evolution necessary for obtaining the expression for $L_\text{f}$.
We provide such a description in Methods, while here we write the final state of the combined quantum system of Alice's random bit (A), the signal component seized by Eve (E), and Bob's memory device storing the measurement outcome (B) after the legitimate users discard invalid bits, i.e., conditional to the successful measurement outcome:
\begin{multline}
    \hat{\rho}^{\text{f}}_\text{ABE} = \sum \limits_{b=0,1} \sum\limits_{a=0,1} 
    \frac{1}{2p(\checkmark|a)}
    \int \!d^2 \alpha\,\,  P(\alpha,\sqrt{T_1}\gamma_a, G_1)
    \\\times\int\!d^2\beta\,\,
    \bra{\beta}\hat{E}_b\ket{\beta}\,
    P\left(\beta, \sqrt{(1-r_\text{E})T_2}\alpha,G_2\right)
    \\\times\ket{a}\bra{a}_\text{A} \otimes \ket{b}\bra{b}_\text{B} \otimes \ket{\sqrt{r_\text{E}}\alpha}\bra{\sqrt{r_\text{E}}\alpha}_\text{E},
    \label{final}
\end{multline}
where 
\begin{equation}
    P(\alpha,\gamma,G) = \frac{1}{\pi (G-1)} \exp\left(-\frac{|\alpha- \sqrt{G} \gamma|^2}{G-1}\right),
    \label{P}
\end{equation}
is the $P$-function describing amplification (see Sec. ``Physical loss control and amplification'' and Supplementary Note 3), and integration operations are performed over the complex plane, i.e., $d^2 \alpha\equiv d\text{Re}(\alpha)\,d\text{Im}(\alpha)$,
$T_{1(2)}$ and $G_{1(2)}$ are, respectively, the transmission probability and amplification factor (the ratio of the output photon number to the input one of an amplification channel) of the effective loss and amplification channels  equivalent to the cascade of amplifiers and losses before (after) Eve's beam splitter (these values depend on the distances between Alice and Eve, $D_\text{AE}$, between Alice and Bob, $D_\text{AB}$, and between neighboring amplifiers, $d$, see Eqs.\,(58--60) in Supplementary Note 3), $p(\checkmark|a)$ is the probability of the successful measurement outcomes in the case that Alice sends bit $a=\{0,1\}$ (the explicit form is given by Eqs.\,(\ref{prob1}) and (\ref{check}) in Methods).

In the case of photon number encoding, Alice randomizes the phase of each pulse.
As a result, neither Bob nor Eve would know the phase $\varphi$ of the incident pulse $\ket{\gamma_a}=\ket{|\gamma_a|e^{i \varphi}}$ which effectively means that the final state of the combined system is described by $\hat{\rho}^{\text{f}}_\text{ABE}$ from Eq.\,(\ref{final}) averaged over $\varphi$ (see Supplementary Note\,4 for details):
\begin{multline}
    \left\langle\hat{\rho}_\text{ABE}^{\,f}\right\rangle_\varphi
    =
    \sum \limits_{b=0,1}
    \sum\limits_{a=0,1}
    \frac{1}{2p(\checkmark|a)}
    \frac{1}{2\pi} \ket{a} \bra{a}_\text{A}
    \otimes
    \ket{b}\bra{b}_{\text{B}}\\\otimes\int\limits_{0}^{2\pi}\!\!d\varphi \cdot \int\!\!d^2\alpha\,\, P(\alpha,\sqrt{T_1}|\gamma_a|e^{i\varphi},G_1)\left|\sqrt{r_\text{E}}\alpha\right\rangle \left\langle  \sqrt{r_\text{E}}\alpha \right| _\text{E}
    \\\times\int\!\!d^2\beta\,\,P\left( \beta,\sqrt{(1-r_\text{E})T_2} \alpha, G_2 \right)
    \braket{\beta|\hat{E}_b|\beta}
    .
    \label{average}
\end{multline}
After the invalid bits are discarded, the information available to Eve about the bits kept by Alice (per bit) is given by
\begin{equation}
\label{iae}
    I(\text{A},\text{E}) = S(\text{A})-S(\text{A}|\text{E}),
\end{equation}
where $S(\text{X})=-\tr\left[\hat\rho_\text{X} \log_2 \hat\rho_\text{X}\right]$ is the quantum von Neumann entropy of system X (which is A, B, E, or their combinations, the corresponding density matrices are obtained from Eq.\,(\ref{final}), or Eq.\,(\ref{average}) if there is phase randomization, by taking partial traces),
and $S(\text{Y}|\text{X})=S(\text{XY})-S(\text{X})$ is the conditional entropy.
We calculate the upper bound of $I(\text{A},\text{E})$ differently in the cases of photon number and phase encoding. 
In the first case, we use the Holevo bound\,\cite{Holevo}, see Supplementary Note\,4.
In the second case, we also rely on the concavity of relative entropy, see Supplementary Note\,5.
\begin{figure*}[t]
\centering
\includegraphics[scale=0.2]{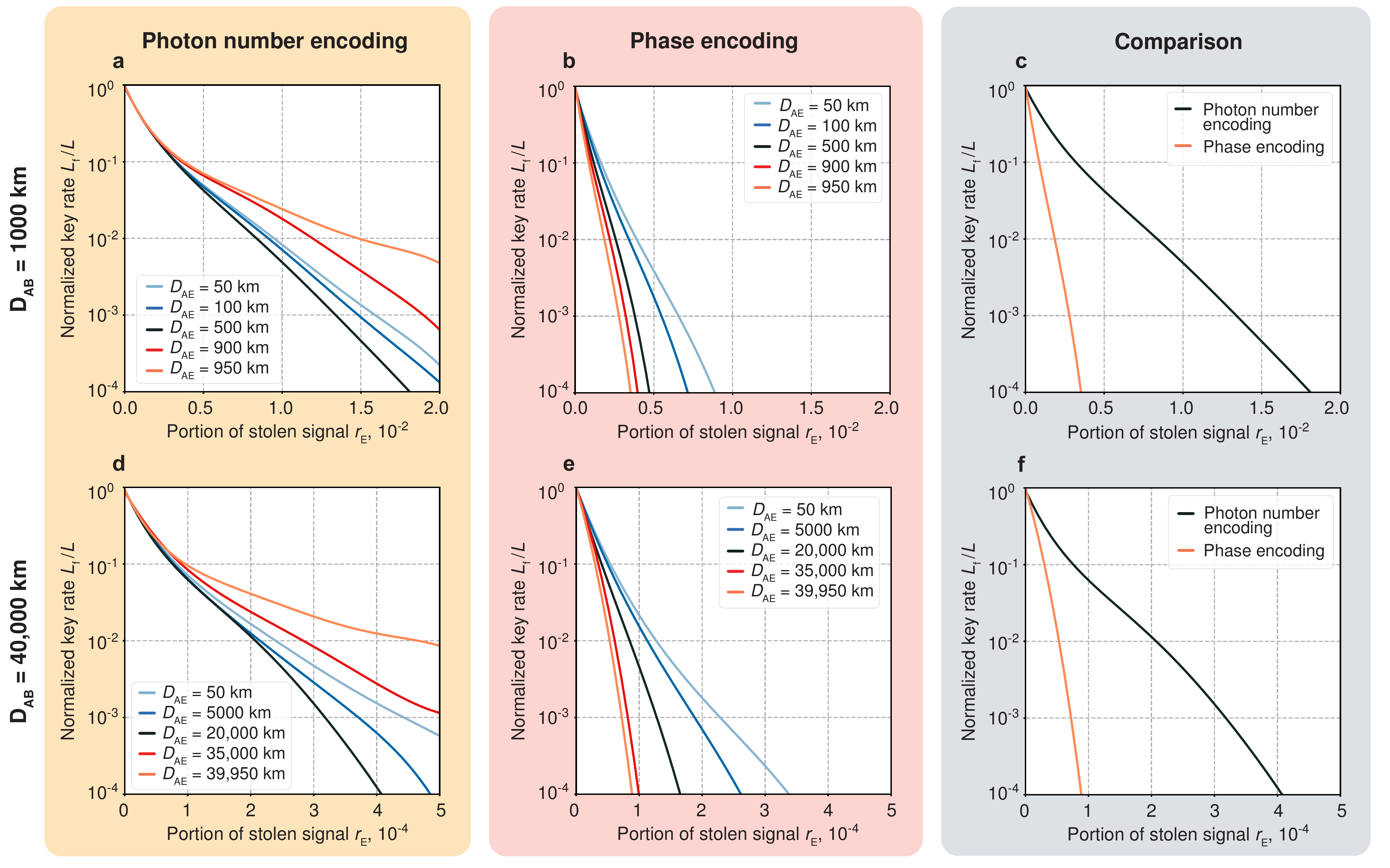}
\caption{\textbf{Numerical simulations of the protocol for different parameters and encoding schemes.}
    \textbf{a} The normalized key rate $L_\text{f}/L$ as function of the proportion of stolen signal $r_\text{E}$ for the photon number encoding and $D_\text{AB}=1000\,\text{km}$. 
    \textbf{b} The same for the phase encoding. 
    \textbf{c} Comparison of the photon number and phase encoding schemes for $D_\text{AB}=1000\,\text{km}$.  \textbf{d} $L_\text{f}/L(r_\text{E})$ for the photon number encoding and $D_\text{AB}=40,000\,\text{km}$. 
    \textbf{e} The same for the phase encoding. 
    \textbf{f} Comparison for the distance $D_\text{AB}=40,000\,\text{km}$.
    In all plots, the distance between neighboring amplifiers $d=50$\,km.
    Different curves in each plot correspond to varying distances between Alice and Eve, $D_\text{AE}$.
    The dependence of $L_\text{f}/L$ on $D_\text{AE}$ is due to the fact that the amount of eavesdropped information is affected by correlations and noise imposed by optical amplifiers.
    The comparative plots \textbf{c,\,f} of two encoding schemes imply the respective worst conditions (with Eve positioned in her best way).
    In each point of every plot, the protocol's parameters---i.e., the photon numbers $\left|\gamma_{0,1}\right|^2$ and postselection parameters $\theta_{1-4}$ (or $\theta_{\text{1,2}}^\prime$)---are numerically optimized for the fixed values of $D_{\text{AB}}$, $D_{\text{AE}}$ and $r_\text{E}$ with respect to $L_\text{f}/L$.
    Depending on $r_\text{E}$, the optimal photon numbers $|\gamma_0|^2$ and $|\gamma_1|^2$ vary from $0.8\cdot10^4$ to $3.0\cdot10^4$ photons in \textbf{a}, from $0.4\cdot10^3$ to $7.2\cdot10^3$ photons in \textbf{b}, from $2.0\cdot10^5$ to $3.0\cdot10^5$ photons in \textbf{d}, and from $0.4\cdot10^5$ to $3.0\cdot10^5$ photons in \textbf{e}.}
\label{Fig5}
\end{figure*}

By performing the error correction procedure, the legitimate users establish a shared bit sequence (raw key) at the price of disclosing an additional error syndrome of the length $f\cdot S(\text{A}|\text{B})$, where $f \geq 1$ depends on the particular error correction code.
As we do not intend to address any specific error correction method, we put $f=1$ corresponding to Shannon's limit.
After the procedure, Eve's information becomes $\tilde I(\text{A},\text{E}) = I(\text{A},\text{E})+S(\text{A}|\text{B})$.
To eradicate Eve's information about the raw key, Alice and Bob perform the privacy amplification procedure tailored precisely for the estimated information leakage due to the local line losses and error correction, see Methods.
The length of the final key is
\begin{multline}
\label{speed}
    L_\text{f}= p_\checkmark L\cdot\left(S(\text{A}) - \tilde I(\text{A},\text{E})\right)
    \\=
    p_\checkmark L\cdot\left(S(\text{A}) - S(\text{A}|\text{B})- I(\text{A},\text{E})\right),
\end{multline}
where $L$ is the number of originally generated random bits, and $p_\checkmark  =\frac{1}{2}\sum_{a,b=0,1} p(b|a)$ is the proportion of bits that are not discarded at the postselection stage.

Taking $L$ and $L_\text{f}$ as the numbers of bits per unit of time, Eq.\,(\ref{speed})---the explicit form of which can be obtained using Eqs.\,(\ref{Eint} or \ref{Ephase}), (\ref{final} or \ref{average}), and Eqs.\,(\ref{prob1},\,\ref{check}) from Methods---gives us the key distribution rate (or key rate for short) as a function of $r_\text{E}$, $|\gamma_0|$, $|\gamma_1|$ and $\theta_{1-4}$ (or $\theta_{\text{1,2}}^\prime$).
Implicitly, the equation also includes the distance between two neighboring amplifiers $d$ and the distances between Alice and Bob, $D_{\text{AB}}$, and Alice and Eve, $D_{\text{AE}}$.
As we specified above, the users obtain the optimal values of $|\gamma_0|$, $|\gamma_1|$ and $\theta_{1-4}$ (or $\theta_{\text{1,2}}^\prime$) by maximizing this analytic formula for measured values of $r_\text{E}$ and $D_{\text{AE}}$.
To be able to distribute secret keys, the users have to possess an information advantage over Eve\,\cite{Devetak_Winter}, which in our case is indicated by the positivity of the calculated $L_\text{f}/L$.
If the evaluated $L_\text{f}/L$ is not positive, the protocol should be terminated.

\section*{Numerical simulations}
Figure\,\ref{Fig5} displays the results of our numerical simulations. 
We plot the optimal normalized key rate $L_\text{f}/L$ as a function of the proportion $r_\text{E}$ for two different transmission distances: \textbf{a,\,b,\,c} $D_{\text{AB}}=1000$\,km, \textbf{d,\,e,\,f} $D_{\text{AB}}=40,000$\,km.
The distance between the neighboring amplifiers $d=50$\,km.

Plots\,\textbf{a,\,d} relate to the photon number encoding with different curves corresponding to different values of $D_{\text{AE}}$.
We observe that the worst normalized key rate $L_f/L$ occurs when Eve is close to the middle of the transmission line.
This is explained by the side effects of signal amplification: the closer Eve is to Bob, the more Eve's part of the signal is correlated with Bob's, yet, the noisier it becomes (see Supplementary Note\,4).
With such a trade-off, Eve gets the largest amount of information, standing somewhere in the vicinity of the line's midpoint.
However, in the phase encoding case, shown in \textbf{b,\,e} panels, correlations outweigh noise even when Eve is close to Bob, resulting in a lower key rate for larger $D_{\text{AB}}$.
Plots\,\textbf{c,\,f} show the protocol's performance under both encoding schemes in the respective worst-case scenarios: Eve's position is such that the key rate is the lowest.
Therefore, as we see from the plots, the photon number encoding scheme appears to be more efficient.

As we show in Methods, see Supplementary Note 3 for technical details, the minimal detectable leakage for a long line with $M$ equidistant amplifiers is $r_\text{E}^\text{min}\sim\sqrt{MG/n}$, where $G$ is the amplification factor of a single amplifier, and $n$ is the number of photons in a test pulse.
With $d=50$\,km, $G=10$, and $n=10^{14}$, we get $r_\text{E}^\text{min}\sim10^{-6}$ and $r_\text{E}^\text{min}\sim10^{-5}$ for the 1000\,km ($M=20$) and $40,000$\,km ($M=800$) lines, respectively.
Close to the loss control precision limit, both encoding schemes allow for high key rates.
For the photon number encoding, the maximum $L_\text{f}/L$ is 0.99 for 1000\,km and 0.57 for $40,000$\,km. 
For the phase encoding, the values are 0.98 and 0.27, respectively.

Within the selected ranges of $r_\text{E}$, which are above the minimum detectable leakage, and, therefore, such losses are resolvable by the physical loss control, and for the photon number encoding we have $L_\text{f}/L\gtrsim 10^{-4}$.
Correspondingly, if the initial random number generation rate $L=1$\,Gbit/s, then for 1000\,km we have $0.99\,\text{Gbit/s}\gtrsim L_\text{f}\gtrsim 100\,$Kbit/s and for 40,000\,km we have $0.57\,\text{Gbit/s}\gtrsim L_\text{f}\gtrsim 100\,$Kbit/s.
In comparison, the asymptotic behavior of the normalized key rate provided by PLOB at a 1000\,km distance is limited to values around $10^{-9}$, or 1\,bit/s for $L=1$\,Gbit/s, which is several orders of magnitude lower than the rates achievable with our method.
To the best of our knowledge, there have been no previously reported QKD protocols that cover tens of thousands of kilometers without using trusted nodes. 
Furthermore, state-of-the-art Twin-Field QKD realizations\,\cite{Toshiba, TF,TF2023} offer key rates that do not exceed a few bits per second at distances comparable to 1000\,km.
At the same time, the QKD realizations featuring high secret key rates of the order of 1\,Kbit/s span relatively short communication distances of a couple of hundred kilometers\,\cite{MDI2016, BB84_2018,TF_2019}.

\section*{Physical loss control and amplification}
Now we outline possible implementations of the basic technological components of the protocol, the physical loss control, and the signal repetition by optical amplifiers.
The physical loss control methods are based on analyzing scattered components of the high-energy test pulses sent along the fiber.
The optical time-domain reflectometry comprises the injection of test pulses into the fiber and subsequent measurement of the temporal sequence of their back-scattered components. 
The response delay defines the distance to a particular scattering point, while its magnitude reflects the respective losses.
Moreover, characteristic features of the response allow for determining the nature of the detected line discontinuity, see the exemplary experimental reflectogram in Fig.\,\ref{reflectogram_exp} in Methods.

\begin{figure}[t]
\centering
\includegraphics[width=0.97\columnwidth]{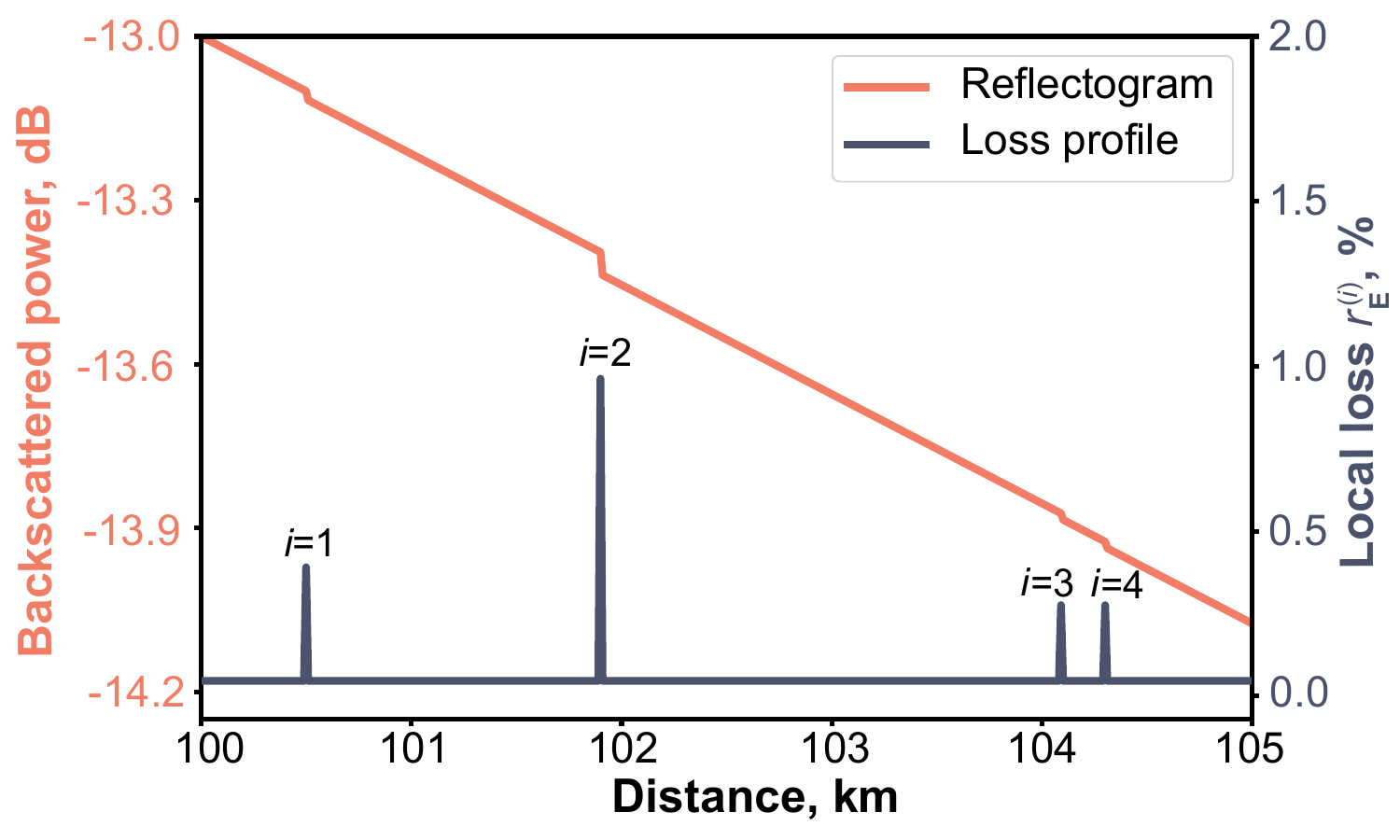}
\caption{\textbf{Exemplary reflectogram and loss profile}. 
The loss profile, which displays the magnitude $r_\text{E}^{(i)}$ of the $i$-th local leakage and its position, is derived from the reflectogram using $\ell_1$-filtering.}
\label{reflectogram}
\end{figure}

As illustrated in Fig.\,\ref{reflectogram}, a log–linear reflectogram features a sequence of linear drops and steep drops (upper trace) corresponding to different discontinuities. 
To construct a loss profile using this piecewise linear graph, one can employ the $\ell_1$-filtering technique\,\cite{l1_filtering}, which is commonly used in the processing of reflectometry data\,\cite{otdr_filt_1,otdr_filt_2}.
This approach involves fitting the graph with a sum of a single linear decreasing function and a series of weighted step-like functions by minimizing the objective function based on the $\ell_1$ norm. 
The $i$-th step-like function's drop (discrete derivative) and its position reveal the corresponding local leak magnitude $r_\text{E}^{(i)}$ and its respective location (lower trace).
In turn, the linear decreasing component defines the homogeneously spread natural scattering losses.

Note, that to obtain an accurate reflectogram one has to make averaging over multiple test runs during which a test pulse travels to the end of the fiber and all its reflections return back.
Accumulating sufficient statistics may, in reality, take a few seconds.
To resolve this problem and to ensure high operational control speed, we develop the component of tomography which we call transmittometry: Alice sends test pulses comprising a large number of photons to Bob, they cross-check the sent and received photon numbers and obtain the proportion of the transmitted photons $(1-r_0)(1-r_\text{E})$.
Knowing the natural losses baseline $r_0$, which is determined during the protocol's initial preparation, they infer the the effective overall leakage $r_\text{E}$ as given by Eq.\,(\ref{total_leak}).
In the spirit of the lock-in method\,\cite{lock-in1994,lock_in1994}, the power of a test pulse can be modulated at high frequency so that the pulse contains a large number of periods.
By comparing the input and output spectral power peaks at the modulation frequency, which can be determined through the Fourier transform of the time-dependent transmitted and received powers, the users can deduce the proportion of losses.
Modulating the power helps to suppress the unwanted contribution of classical noise that may be present in the system. 
As the modulation frequency increases, the corresponding value of the noise spectral density tends to decrease, raising the transmittometry precision.
Unlike reflectometry, transmittometry does not enable the users to localize and identify individual leakages but immediately updates the estimate of the effective overall leakage.
Thus, the two control methods complement each other, the users are constantly aware of the magnitude of leakages and can localize them after accumulating sufficient reflectometry statistics.

To discriminate between the intrinsic and artificial line losses, the legitimate users prerecord the initial undisturbed line tomogram, including the reflectogram and the total proportion of losses in the line, and use this tomogram as a reference.
The fiber material, silica, has an amorphous nonreproducible structure, making its reflectogram a physically unclonable function\,\cite{fiberPUF}. 
With that, the fiber core can be slightly doped, with, e.g., Al, P, N, or Ge, to tune its tomography results and achieve the optimal parameters such as dispersion.
The most general eavesdropping attack implies a unitary transformation of the state of the combined system comprising the propagating signal and some ancillary eavesdropping system.
However, coupling of photons to devices outside the line requires making significant alterations to the fiber medium, which would inevitably change the reflectogram and hence will be detected.
Quantum cryptography also addresses attacks exercising the partial blocking of the signal and the subsequent unauthorized substitution of the blocked part.
Any intervention like that would inevitably and permanently (even if Eve at some point decided to disconnect from the line) affect the tomogram of the transmission line and hence will be detected by the legitimate users.

The key distribution itself should go in parallel with accumulating the reflectometry statistics. 
If, at some point, the reflectogram shows an intrusion into the line, the users should respond with the appropriate postprocessing of the bits distributed during the formation of the reflectogram.
This may possibly come down to discarding the whole bit sequence.
Ideally, the physical loss control should be conducted permanently and should not halt even during the pauses in the key distribution. 
Taking the transmittometry test pulses' duration of the order of 1\,ns makes any real-time mechanical intrusion into the line immediately detectable.
\begin{figure*}[t]
\centering
\includegraphics[scale=1.1]{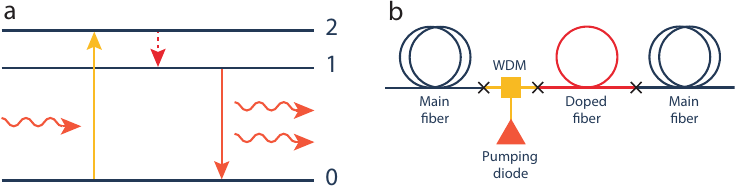}
\caption{\textbf{Optical amplification.} \textbf{a} Energy diagram of the light amplification in the erbium-doped fiber section.
Pumping radiation excites erbium ions from the ground state 0 into the 2nd energy level. 
Shortly after, ions drop to the metastable level 1. 
The incident photons stimulate the transition $1\rightarrow0$ which results in the coherently synchronized radiation of additional photons at the same wavelength.
\textbf{b} Schematics of the proposed bidirectional optical amplifier.
The doped fiber section is embedded into the main fiber line and linked to the pumping diode through the WDM.}
\label{Fig2}
\end{figure*}

The principal task of the physical loss control is to ensure that Eve does not get enough photons to obtain the informational advantage over Bob.
With that, signal pulses and test pulses still carry large numbers of photons, making it possible to repeat them via optical amplification.
The repeater can particularly be arranged as a doped fiber section embedded into the main line and pumped to produce amplification gain in the primary mode.
In telecommunications, the most common dopant is erbium. 
Pumped at the wavelength of 980\,nm, the erbium-doped fiber generates the gain at around 1550\,nm which fits into the transmission window of the silica-based fiber.
Upon absorption of the pumping radiation, erbium ions transit from the ground state (state 0) to a short-lived state (state 2). 
From there, they non-radiatively relax to a metastable state (state 1), as illustrated in Fig.\,\ref{Fig2}\textbf{a}. 
As the signal photonic mode passes through the inverted atomic medium, it stimulates the transition from state 1 to state 0, resulting in a coherently synchronized photon emission.
The resulting signal amplification magnitude depends on the erbium ion concentration, the length of the doped fiber segment (active fiber), and the power of the pumping radiation. 
The interaction between the propagating signal photonic mode (with the annihilation operator $\hat{a}$) and the inverted atoms is determined by the Hamiltonian in the rotating wave approximation
\begin{equation}
    \hat{H}_{\text{int}} = i\kappa
    \sum\limits_{n=1}^{N}\left(
    \hat{a}^\dag \otimes\ket{0}\bra{1}_{n}
    -
    \hat{a}\otimes\ket{1}\bra{0}_{n}
    \right),
\end{equation}
where atoms are indexed by $n$, $N\gg 1$ is the total number of atoms in the medium, $\kappa$ is the interaction constant, and states $\ket{0}_n$ and $\ket{1}_n$ represent respectively states 0 and 1 of the $n$-th ion.
Since the time of relaxation from state 2 to 1 is very small (20\,$\mu$s against 10\,ms for the relaxation from 1 to 0), we take that an ion is effectively a two level system.
The Holstein-Primakoff transformation maps the active medium's state with $m$ atoms in state 0 and $N-m$ in state 1 to a Fock state $\ket{m}_b$.
This state corresponds to $m$ excitations in the auxiliary bosonic mode associated with the annihilation operator $\hat{b}$.
With that, we get the following equation
\begin{equation}
    \hat{H}_{\text{int}}
    =
    i\kappa \sqrt{N}\left(\hat{a}^{\dagger}\hat{b}^{\dagger}
    -
    \hat{a}\hat{b}
    \right).
\end{equation}
The evolution operator describing the signal mode propagation is given by
\begin{equation}
\label{U_g}
    \hat{U}_g = e^{-i\hat{H}_{\text{int}} t/\hbar}=e^{g (\hat{a}^\dag \hat{b}^\dag - \hat{a} \, \hat{b})},
\end{equation}
where $g = \kappa\sqrt{N}t/\hbar$, and $t$ is the effective time of interaction between the signal mode and active medium.
Amplification is described by a quantum channel acting on the signal's density matrix $\hat{\rho}$
\begin{equation}
\label{channel}
    \text{Amp}_{G=\cosh^2(g)}[\hat{\rho}]=\tr_b \left[ \hat{U}_g\,   \hat{\rho} \otimes \ket{0} \bra{0}_b\, \hat{U}_g^\dag \right],
\end{equation}
where the partial trace is taken over the states of the auxiliary mode.
Within the $P$-function formalism, Eq.\,(\ref{channel}) translates into Eq.\,(\ref{P}). 
Additional details can be found in Supplementary Note 3.

Usually, doped fiber amplifiers utilize optical isolators, which allow the light to pass only in one direction. 
This minimizes the risk of multiple reflections inside the doped fiber section.
In our protocol, however, the optical isolators would block the reflected light hindering the end-to-end time-domain reflectometry.
Besides, the amplifiers typically include tap couplers diverting about $1\% $ of the radiation into the photodetectors to monitor the amplifiers' operation, and this fraction can possibly be seized by the eavesdropper.
We hence opt out of both the optical isolators and tap couplers and utilize the design of the bidirectional optical amplifier \cite{BidirectEDFA2012, BidirectEDFA2013, BidirectEDFAQKD}.
The amplifier's sketch is displayed in Fig.\,\ref{Fig2}\textbf{b}. 
The fiber core is connected to the wavelength-division multiplexing (WDM) system.
The WDM system is a beam splitter-like device for guiding the radiation of the different wavelengths into a single optical fiber. 
In our case, it is intended to feed the doped fiber section with the pumping radiation necessary to excite the active fiber's dopant atoms.
Correspondingly, the WDM is connected to the active fiber and the pumping diode.
Finally, active fiber is connected to the main fiber line.
Provided that the neighboring amplifiers are separated enough, they are not subject to significant cross-talk. 

Our preliminary experiments reveal how the 1000\,km-long line with the standard telecom distance $d=50$\,km between amplifiers can be made stable with the very restricted signal noise in the line, even in the absence of optical isolators.
The stability and scalability of our QKD scheme are supported by the remarkable precision of the loss control observed in the experiments.
This control precision is accomplished through the use of high-resolution reflectometry, complemented by optical amplifiers to extend its range, and lock-in based transmittometry.
With the ability to capture leakages above the control resolution, the users maintain an information advantage over potential adversaries attempting to exploit these leakages, as confirmed by analytical calculations. 
Coupled with the experimentally observed low bit error rates, our QKD system ensures high rates of secure key generation.
We find that the signal wavelength of 1530\,nm---corresponding to the peak of the amplification factor spectrum of the erbium-doped fiber amplifier---is more preferable than the standard 1550\,nm wavelength.
Fixing $G=1/T$ for 1550\,nm means greater amplification for the noise in the modes near 1530\,nm, which, in turn, may disrupt the stability of the amplifiers' operation, possibly turning them into lasers.
But this is not the case if 1530\,nm is already the target wavelength itself.

A possible eavesdropping attack on the amplifier may consist of increasing the pumping power and stealing the surplus of the amplified radiation.
Of course, hooking up to the line would change its tomogram and thus will be detected.
Nevertheless, the following constructive feature of the amplifier will serve as an additional element of protection. 
The doped fiber section will contain the near minimum number of dopant ions necessary to amplify the signal with the target amplification factor. 
Let the operational pumping power $P_\text{p}$ match the target amplification factor $G$.
With the increase of the pumping power, the relative population inversion asymptotically approaches unity, which corresponds to the amplification factor $G+\delta G$.
The fraction of signal that Eve can possibly steal by inflating the pumping power is limited by $\delta G/G$, which is small, if at $P_\text{p}$ almost all of the ions are already excited.
The number of ions and the value of $P_\text{p}$ should be such that
the maximum achievable Eve's makeweight, summed over all amplifiers installed into the line, is smaller than the minimum detectable leakage.

\section*{Discussion and conclusions}
Quantum cryptography typically assumes channel device independence, suggesting that an eavesdropper can fully exploit all leakages from the quantum channel. 
This assumption constrains key rates to the PLOB bound\,\cite{PLOB}, where the maximum rate scales as $-\log_2(1 - T)$ bits per channel with the transmittivity $T$. 
Consequently, key rates become impractically small over long distances.

By examining the physical quantum principles governing signal transmission and implementing physical loss control, we shifted this paradigm. 
In our approach, legitimate users can determine the fraction of losses accessible to an eavesdropper and ensure it contains insufficient information. 
While Eve struggles to discriminate between weak quantum states, Bob receives signals with relatively large numbers of photons, granting him a significant information advantage.
Our approach employs end-to-end line tomography and leverages the impossibility of exploiting natural losses. 
As a result, the PLOB constraint is lifted, significantly extending the practical implementation of QKD over long distances without sacrificing security---notably, without relying on trusted nodes. 
Optical amplifiers, unlike trusted nodes, do not convert quantum information into classical form and are directly controlled by users through end-to-end control.

In this study, we analyzed a physical model restricting Eve to the possibility of local leakage exploitation, demonstrating the security of loss control-based QKD under long-distance transmission and high signal intensity conditions. 
Our forthcoming mathematical research will explore alternative physical models involving more complex actions by Eve and will provide further security proofs, particularly dealing with the finite key length and security parameter\cite{portmann2022security}.

The proposed approach maintains the fundamental advantage of the QKD, everlasting security\,\cite{unruh2013everlasting,portmann2022security,stebila2009case,alleaume2010quantum}, ensuring that distributed keys will remain secure even against future technologies or attacks that may be developed.
In the forthcoming publication, we will address the experimental realization of the QKD based on our approach for the transmission distance over 1000 kilometers.

\section*{Methods}

\begin{figure*}[t]
\centering
    \includegraphics
    [width=\linewidth]{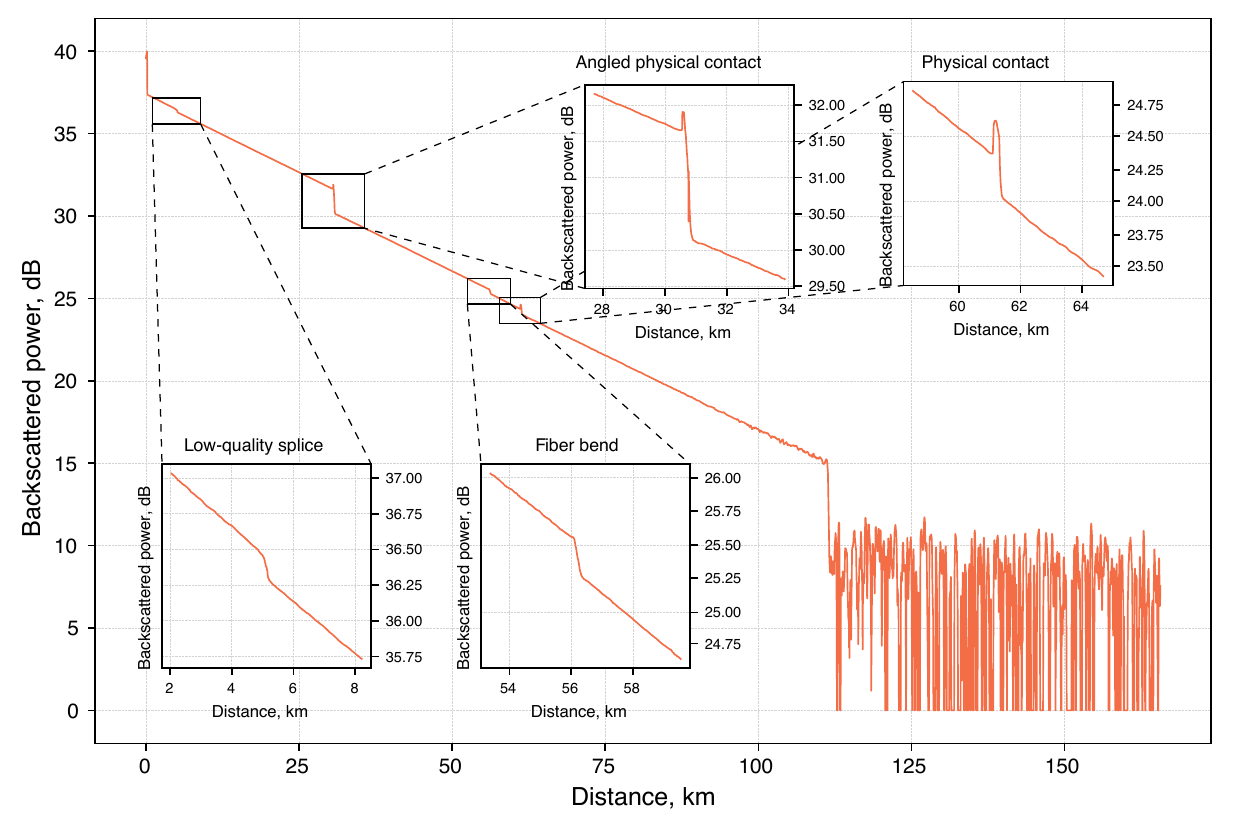} 
    \caption{\textbf{An exemplary plot obtained with the optical time-domain reflectometer.}
    The device sends the high-intensity test pulses into the fiber and registers its reflections providing the dependence of the backscattered power on the distance to the scattering point defined by the time of the signal's return.
    }
    \label{reflectogram_exp}
\end{figure*}

\subsection*{Combined quantum state evolution under beam splitting attack}
Here we provide the description of states' evolution in the case of the beam splitting attack.
We will use that (a) an amplifier transforms pure coherent state into a mixture of the coherent states, 
\begin{equation}
    \ket{\gamma}\rightarrow \int d^2\alpha\,P(\alpha,\gamma,G) \ket{\alpha}\bra{\alpha},
\label{amp}
\end{equation}
where $P(\alpha,\gamma,G)$ is given by Eq.\,(\ref{P})
and integration is performed over the complex plane with $d^2 \alpha\equiv d\text{Re}(\alpha)\,d\text{Im}(\alpha)$; and that
(b) formally, a sequence of losses and amplifications can be reduced to a single pair of the loss and amplification quantum channels---see Supplementary Note 3 for details.

The initial density matrix of Alice's random bit (A) and the corresponding signal (S) is given by
\begin{equation}
    \hat{\rho}^\text{i}_\text{AS}= \frac{1}{2}\ket{0}\bra{0}_\text{A} \otimes \ket{\gamma_0}\bra{\gamma_0}_\text{S} +
    \frac{1}{2}\ket{1}\bra{1}_\text{A} \otimes \ket{\gamma_1}\bra{\gamma_1}_\text{S}.
    \label{initial_state}
\end{equation}
Just before the signal passes the beam splitter, the state of the AS system is
\begin{multline}
    \hat\rho^{\rightarrow \Box}_\text{AS} = \frac{1}{2} \sum \limits_{a=0,1} \ket{a} \bra{a}_\text{A} \\\otimes \int d^2 \alpha \cdot P(\alpha,\sqrt{T_1} \gamma_a,G_1) \cdot  \ket{\alpha} \bra{\alpha}_{\text{S}},
\end{multline}
where we use Eq.\,(\ref{amp}) to describe the state of sequentially attenuated and amplified signal, and $T_1$ and $G_1$ are, respectively, the transmission probability and amplification factor of the effective loss and amplification channels that are equivalent to the sequence of amplifications and losses prior to the beam splitter, see Supplementary Note 3, particularly Eq.\,(58).
Just after the signal passes the beam splitter, the state of the joint system of Alice's random bit, the signal travelling to Bob and the signal component seized by Eve (E) is described by
\begin{multline}
    \hat{\rho}^{\Box \rightarrow}_\text{ASE} = \frac{1}{2} \sum \limits_{a=0,1} \ket{a} \bra{a}_\text{A} \otimes \int d^2 \alpha \cdot P(\alpha,\sqrt{T_1} \gamma_a,G_1) \\\times \ket{\sqrt{1-r_\text{E}} \alpha} \bra{\sqrt{1-r_\text{E}} \alpha}_\text{S} \otimes \ket{\sqrt{r_\text{E}} \alpha} \bra{\sqrt{r_\text{E}} \alpha}_\text{E},
\end{multline}
where $r_\text{E}$ is the fraction of signal stolen by Eve.
After the signal passes the second series of losses and amplifiers and right before it is measured by Bob, the state of the joint system is
\begin{multline}
\label{toBob}
    \hat{\rho}^{\rightarrow \text{Bob}}_\text{ASE} = \frac{1}{2} \sum \limits_{a=0,1} \ket{a} \bra{a}_\text{A} \otimes \int d^2 \alpha \cdot P(\alpha,\sqrt{T_1} \gamma_a,G_1) \\\times \left( \int d^2 \beta \cdot P\left( \beta,\sqrt{(1-r_\text{E})T_2} \alpha, G_2 \right) \cdot \ket{\beta} \bra{\beta}_\text{S} \right)  \\\otimes \ket{\sqrt{r_\text{E}} \alpha} \bra{\sqrt{r_\text{E}} \alpha}_\text{E},
\end{multline}
where we again utilize Eq.\,(\ref{amp}) to describe the evolved signal state, and $T_2$ and $G_2$ are the effective transmission probability and amplification factor of the region between the beam splitter and Bob, see Eq.\,(60) in Supplementary Note 3.
Bob receives the signal state, measures it and, together with Alice, discards the bits corresponding to the failed measurement results.
The probability that Bob's measurement outcome is $b=\{0,\,1\}$ given that Alice's sent bit is $a=\{0,\,1\}$ can be written as
\begin{equation}
\label{prob1}
p(b|a)=\tr_\text{ASE}\left[{\left(2\cdot\ket{a}\bra{a}_\text{A} \otimes \hat{E}_b\otimes \hat{\mathds{1}}_\text{E}\right) \hat{\rho}^{\rightarrow \text{Bob}}_\text{ASE}}\right],
\end{equation}
where $\hat{E}_b$ is given by Eq.\,(\ref{Eint}) or (\ref{Ephase}) depending on the encoding scheme, and $\tr_\text{ASE}[\dots]$ is the trace over the ASE system.
The probability that Bob performs a successful measurement if Alice sends bit $a$ is 
\begin{equation}
    p(\checkmark|a)=p(0|a)+p(1|a).
    \label{check}
\end{equation}
Equation\,(\ref{final}) follows from Eq.\,(\ref{toBob}) after applying measurement operators to the signal subsystem, discarding sum components associated with the failed outcomes, and renormalizing.

\subsection*{Privacy amplification}
Privacy amplification can be realized through applying the universal hashing method\,\cite{Carter_universal_hashing}, which requires the users to initially agree on the family $\mathcal{H}$ of hash functions.
At the privacy amplification stage\,\cite{Bennett_PA, Bennett_PAgeneralized, Cascade1}, they randomly select such a function $h:\{0,1\}^{p_\checkmark L}\rightarrow \{0,1\}^{L_\text{f}}\in \mathcal{H}$ that maps the raw key of length $p_\checkmark L$ to the final key of length $L_\text{f}$. 
If Eve is estimated to know $p_\checkmark L\cdot\tilde I(\text{A},\text{E})$ bits of the raw key, the letter must be taken in accord with Eq.\,(\ref{speed}).
Family $\mathcal{H}$ can, for example, span Toeplitz matrices\,\cite{krawczyk_Toeplitz}: a random binary Toeplitz matrix $\hat{T}$ with $p_\checkmark L$ rows and $L_\text{f}$ columns translates the binary vector representation of the raw key $\mathbf{v}$ into the vector $\mathbf{k}$ representing the final key, $\mathbf{k} = \hat{T}\cdot \mathbf{v}$.
\subsection*{Physical loss control precision}
Assume that Bob is equipped with an optical filter with a very narrow wavelength band which blocks noise from the secondary light modes (for details on this additional noise, see Supplementary Note 3).
Assume also that all amplifiers are positioned equidistantly, each having amplification factor $G=1/T$ with $T$ being the transmission probability of the line section between two neighboring amplifiers.

Approaching an amplifier, the test pulse comprising $n$ photons is attenuated down to $Tn$ photons. 
The amplifier restores the number of photons back to $n$ but adds noise. 
The photons in the pulse follow the Poisson  statistics; thus, the photon noise just before the amplifier can be taken as a square root of the number of photons in the input signal $\sqrt{Tn}$. 
The noise is amplified by factor $G$ as well, so after a single amplifier the noise is $G \sqrt{T n}$.
Coming through a sequence of $M$ amplifiers which add fluctuations independently, the total noise raises by the factor $\sqrt{M}$.
The noise at Bob's end is thus $\delta n_\text{B}\simeq G \sqrt{MT n}=\sqrt{GM n}$.
The minimum detectable leakage can be calculated as $r_\text{E}^\text{min}\sim\delta n_\text{B}/n=\sqrt{MG/n}$.
Our qualitative estimates match with the detailed calculations in Supplementary Note 3.

\subsection*{Experimentally obtained reflectogram}
Figure\,\ref{reflectogram_exp} displays an example of an experimentally obtained reflectogram. 
Every particular type of fiber discontinuity, whether it is physical contact, bending, or splice, can be identified by its own unique reflectographic pattern, as demonstrated in the inset plots.
The appearance of peaks signifies the excessive scattering which happens, for instance, at the physical connectors where the signal undergoes the Fresnel reflection.
The right noisy tail of the main plot corresponds to the end of the backscattered signal.
The measurements are carried out with a 2\,$\mu$s 1550\,nm pulse laser with a power up to 40\,mW.
The experimental data is averaged over $16,000$ measurements.
\newpage

%%%%%%%%%%%%%%%%%%%%%%%%%%%%%%%%%%%%%%%%%%%%%%
%%%%%%%%%%%%%%% BIBLIOGRAPHY %%%%%%%%%%%%%%%%%
%%%%%%%%%%%%%%%%%%%%%%%%%%%%%%%%%%%%%%%%%%%%%%

%%%%%%%%%%%%%%%%%%%%%%%%%%%%%%%%%%%%%%%%%%%%%%%%%%%%%
%%%%%%%%%%%%%% SUPPLEMENTARY INFO %%%%%%%%%%%%%%%%%%%
%%%%%%%%%%%%%%%%%%%%%%%%%%%%%%%%%%%%%%%%%%%%%%%%%%%%%

\afterpage{\null\newpage}
\newpage
\section*{\large{SUPPLEMENTARY INFORMATION}}
\section*{NOTE 1. Natural losses}
In this note, we discuss the unfeasibility of eavesdropping on the natural losses occurring due to the scattering of photons in optical fiber. 
Here we explore quantum considerations analogous to those that serve to derive the Second Law of Thermodynamics\,\cite{Lesovik_Qirreversibility, Kirsanov_Qirreversibility, Lesovik2_Qirreversibility}.

In quantum cryptography, the thermodynamic considerations should focus on the collection of optical states traveling through a lossy fiber. 
In the standard telecommunication scenario, with the signal power of around 10\,mW and a pulse duration of 1\,ns, each light pulse contains about $10^8$ photons, with staggering leaks of about $10^3$ photons per meter that can be measured outside of the fiber.
This scenario implies that the system is clearly not isolated, and, through the measurement of leaked photons, Eve can decrease the system's entropy---a process which, in accordance with Shannon's definition of entropy, is equivalent to extracting information.

On the other hand, in a relatively weak signal region of our choice with $10^4$ to $10^5$ photons or less (which is still enough for the optical amplification), the system can be considered quasi-isolated since only a small number of photons are lost from each pulse. 
Eve's attempt to measure the leaked parts of a signal’s wave function and obtain information about the sent bit would decrease the entropy. 
However, extracting a substantial amount of information from leaked photons requires Eve to operate and observe multitudes of degrees of freedom, which, as we will demonstrate in the next section, is unfeasible.

\subsection*{1.1 $\quad$ Required length of the eavesdropping device}
In assessing the feasibility of the potential eavesdropping on the natural losses, we consider a scheme in which logical bits are encoded into optical coherent states $\ket{\gamma_0}$ and $\ket{\gamma_1}$ with different photon numbers $\mu_0=\left|\gamma_0\right|^2$ and $\mu_1=\left|\gamma_1\right|^2$.
Let the signal pulses have the duration of 1\,ns ($0.2$\,m long) and comprise $10^{4}$ photons on average.
The optimal values for $\mu_0$ and $\mu_1$ on a 1000 km line, as determined by our simulations (see the main text), are 9000 and 11,000, respectively. 
To eavesdrop on the homogeneously spread natural losses, an eavesdropper would be forced to undertake measurements along various segments of the fiber, possibly using single-photon detectors. 
However, as our calculations indicate, such a method would be impractical in terms of the sheer length required to successfully determine the value of any given bit.

The natural losses coefficient of a fiber section of length $l$ can be calculated as
\begin{equation}
    r_l=1-10^{-\xi\cdot l},
\end{equation}
where $\xi$ is the decay constant. 
The number of photons lost from a wave packet containing $\mu_a$ photons is given by:
\begin{equation}
    \mu_\text{E}^{(a)}=\mu_a\cdot r_l.
\end{equation}
The lower index E denotes that the photons can be seized by an eavesdropper, the upper index $a$ represents the corresponding random bit value.

The observable (positive operator-valued measure, or POVM) describing the single-photon detector includes two projective operators corresponding to two possible outcomes: 
\begin{equation}
    \mathcal{M}_\text{s.ph.}
    =
    \left\{\hat{M}_0=\ket{0}\bra{0},\,\hat{M}_{\text{click}}=\sum\limits_{n=1}^{+\infty}\ket{n}\bra{n}\right\}.
\end{equation}
The probability of the detector's ``click'' conditional to the bit value $a$ is determined by the Poisson statistics of Eve's coherent state
\begin{multline}
    q_a\equiv p\left(\text{click}\,|\,a\right)
    =
    \text{Tr}\left(\hat{M}_{\text{click}}\cdot\ket{\sqrt{r_l}\gamma_a}\bra{\sqrt{r_l}\gamma_a}\right)
    \\=
    1-\text{Tr}\left(\hat{M}_0\cdot\ket{\sqrt{r_l}\gamma_a}\bra{\sqrt{r_l}\gamma_a}\right)
    =
    1-\left|\braket{0|\sqrt{r_l}\gamma_a}\right|^2
    \\=
    1-e^{-|\sqrt{r_l}\gamma_a|^2}
    =
    1-e^{-\mu_\text{E}^{(a)}}.
\end{multline}
According to the measurement outcomes, Eve makes bit decisions.
Probability distribution of measurement results can be considered as Binomial which variance is $q_a(1-q_a)$.
Carrying out $N$ independent measurements of sequential parts of the line, the combined variance is a sum of variances of each individual measurement.
Thus, the expression for the square root of the variance takes form
\begin{equation}
    \delta n_a = \sqrt{N \cdot q_a(1-q_a)}.
\end{equation}
Bits zero and one produce different distributions, the distance between the maximums of these distributions can be calculated as
\begin{equation}
    \Delta n = N \cdot \left|\mu_\text{E}^{(1)}-\mu_\text{E}^{(0)}\right|.
\end{equation}
In order to obtain significant amount of information Eve needs the distance between the maximums of the distributions to exceed the sum of their standard derivations, i.e. the notional critical condition can be written as 
\begin{multline}
    \Delta n=\delta n_0+\delta n_1
    \quad\Rightarrow\\
    \!\!N\left|\mu_\text{E}^{(1)}-\mu_\text{E}^{(0)}\right|
    \!=\!
    \sqrt{N}\!\left(\sqrt{q_0(1-q_0)}\!+\!\sqrt{q_1(1-q_1)}\right)\!.
\end{multline}
Then, supposing that each of the detectors covers a piece of fiber of the length equal to the length of the considered pulses, i.e. $l=0.2$\,m, and taking $\xi=0.02\,\text{km}^{-1}$, which is a common value for the single-mode optical fiber, the required number of detectors
\begin{equation}
    N = 
    \frac{\left(\sqrt{q_0(1-q_0)}+\sqrt{q_1(1-q_1)}\right)^2}{\left|\mu_\text{E}^{(1)}-\mu_\text{E}^{(0)}\right|^2}
    \simeq 10^3.
\end{equation}
Combining all measured pieces, we obtain the total length of the whole detection device $N\cdot l\simeq10^3\cdot 0.2\,\text{m}=200$\,m.

 Moreover, not all the leakages can be measured in reality: some of the scattered photons transform to different modes propagating along the fiber and do not radiate outwards.
 Along with that, to measure a leaked part of the signal with the single photon detector, it is necessary to isolate the measured part of the line from the external radiation and concentrate the leaked photons to the cryogenic setup.
 The effects combined reduce the number of photons available to Eve approximately by an order of magnitude.
 These physically motivated assumptions lead to enormous lengths of detection devices even in the case of higher intensities (for instance, $\mu_{0,1}\sim 10^5$, which turned out to be optimal for the 40\,000\,km line -- see Numerical Simulations).
It is important to acknowledge that while the aforementioned attacks utilizing scattering losses are hardly realistic, one can employ methods of protection that can counteract them as well. 
These include utilizing specialized cable design enabling the controlled dissipation of scattering losses (described in Note 2), utilizing lower numbers of photons in pulses, and implementing advanced encoding schemes that utilize the phase or shape of the pulses.

%\subsection{Eve's information estimation}
\subsection*{1.2 $\quad$ Precise estimation of Eve's information}
To estimate the precise amount of information that Eve can get from the $N$ individual measurements of natural losses with single photon detectors, we calculate the mutual information between Alice's sent bit $a\in \{ 0,1\}$ and Eve's measurement results.
The conditional probabilities of obtaining $n$ ``clicks'' can be written as
\begin{equation}
\begin{gathered}
    p(n|a) = C_N^n \cdot q_a^n \cdot (1-q_a)^{N-n},
    \label{conditional_probabilities}
\end{gathered}
\end{equation}
where $C_N^n=\frac{N!}{n!(N-n)!}$ is a binomial coefficient and $q_a$ is the click probability in the individual measurement in the case where the sent bit value is $a$.
The expression for the mutual information is determined by the joint probability distribution $p(n,a)=p(n|a)\cdot p(a)$: 
\begin{multline}
    I\left(\text{A}:\text{E}^{(N)}_{\text{ind}}\right)
    =
    \sum\limits_{n=0}^N
    \sum\limits_{a=0}^1
    p(n,a)\cdot\log_2
    \left(\frac{p(n,a)}{p(n)\cdot p(a)}\right)
    \\=
    \sum\limits_{n=0}^N
    \sum\limits_{a=0}^1
    \frac{1}{2}p(n|a)\cdot\log_2\left(\frac{p(n|a)}{p(n|0)+p(n|1)}\right)
    \\=
    1-\frac{1}{2}\sum\limits_{n=0}^N
    \Big(p(n|0)+p(n|1)\Big)
    \\\times h_2\left(\frac{p(n|0)}{p(n|0)+p(n|1)}\right).
    \label{information_individual}
\end{multline}
The dependence of the Eq.\,(\ref{information_individual}) on the total length of the detection device is depicted in Fig.\,\ref{dozens}.
Was the device's length equal to 200\,m, the mutual information $I\left(\text{A}:\text{E}^{(N)}_{\text{ind}}\right)$ would almost reach 0.5, meaning that Eve would know half of the raw shared key---these results are in a good agreement with the preliminary estimations from the previous section.
While in the case of more feasible lengths (a couple of meters, for example), the information Eve can obtain is of the order of $10^{-2}$\,bit.
One can note that the Eq.\,(\ref{information_individual}) does not depend on whether phase randomization is applied or not, since the probabilities are determined only by photon numbers (for details see Note 4). 

\subsection*{1.3 $\quad$ Ideal photon number measurement}
\begin{figure}[t]
\noindent\centering{
\includegraphics[width=0.8\columnwidth]{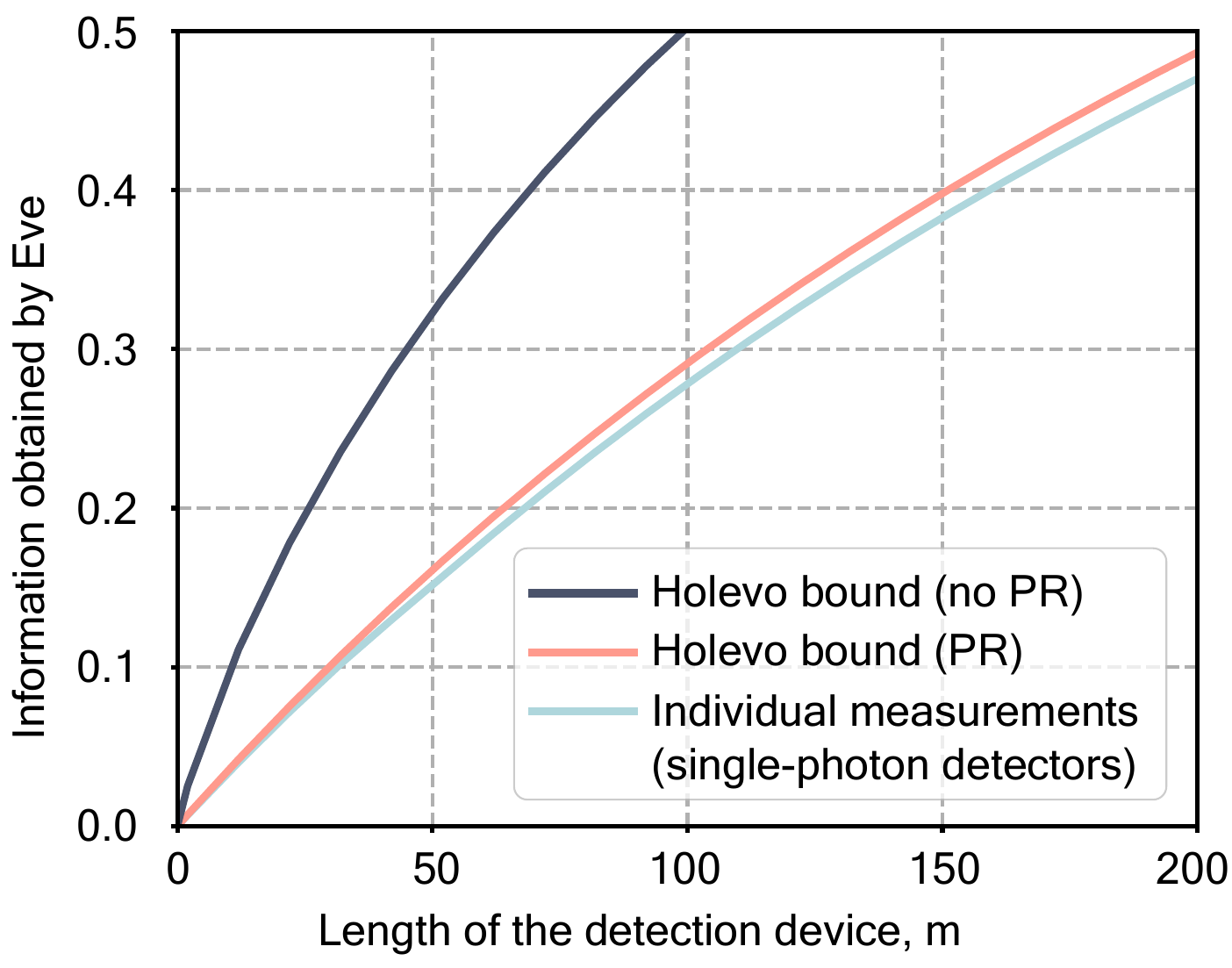}}
\caption{
\textbf{The information Eve can obtain from natural losses as a function of the overall length of the detection device.}
The cyan line corresponds to individual measurements (by single-photon detectors) as determined by Eq.\,(\ref{information_individual}). The orange line depicts Holevo bound with phase randomization (PR) according to Eq.\,(\ref{Holevo_PR}). The black line is for the Holevo bound (no PR) in the absence of PR Eq.\,(\ref{Holevo_No_PR}).}
\label{dozens}
\end{figure}

Next, we consider collective photon number measurement over the whole $N$ pieces of the fiber 
which can be conducted after gathering all the scattered photons at one ideal detector.
The corresponding observable in terms of the POVM effects can be expressed as projectors on the Fock-states
\begin{equation}
    \mathcal{M}_{\text{ photon number}}
    =
    \left\{\hat{M}_n=\ket{n}\bra{n}\right\}_{n=0}^{+\infty}.
\end{equation}
The probability of obtaining $k$ photons is determined by the average number of all scattered photons in a signal corresponding to sent bit $a$
\begin{equation}
    p(k|a) = e^{-N \cdot \mu_\text{E}^{(a)}} \cdot \frac{\left( N\mu_\text{E}^{(a)} \right)^k}{k!}.
    \label{Poisson_probabilities}
\end{equation}
Almost analogously to the previous paragraph, one can calculate the mutual information as 
\begin{multline}
    I\left(\text{A}:\text{E}^{(N)}_\text{col}\right)
    =
    1-\frac{1}{2}\sum\limits_{k=0}^{+\infty}
    \Big(p(k|0)+p(k|1)\Big)\\\times h_2\left(\frac{p(k|0)}{p(k|0)+p(k|1)}\right).
    \label{information_photon_number}
\end{multline}
The only difference from the Eq.\,(\ref{information_individual}) is that upper limit of the summation is now infinity.
As depicted in the Fig.\,\ref{dozens}, the informational advantage of Eq.\,(\ref{information_photon_number}) over single-photon measurements is insignificant.

\subsection*{1.4 $\quad$  The Holevo bound}
To build an upper-bound on the information that Eve may extract from the scattered photons, we calculate the Holevo quantity, which for an ensemble of quantum states $\mathcal{E}=\left\{\left(1/2,\,\hat{\rho}^{(0)}\right),\:\left(1/2,\,\hat{\rho}^{(1)}\right)\right\}$ is defined as 
\begin{multline}
    \chi\left(\mathcal{E}\right)
    =
    S\left(\frac{1}{2}\hat{\rho}^{(0)}+\frac{1}{2}\hat{\rho}^{(1)}\right)
    \\-
    \frac{1}{2}S\left(\hat{\rho}^{(0)}\right)
    -
    \frac{1}{2}S\left(\hat{\rho}^{(1)}\right),
\end{multline}
where $S(\hat{\rho})=-\text{Tr}\left(\hat{\rho}\cdot \log_2\hat{\rho}\right)$ is von Neuman entropy. 
Without phase randomization, Eve has to distinguish between pure coherent states of the form
\begin{equation}
    \hat{\rho}^{(a)} =\ket{\sqrt{r_lN}\gamma_a}\bra{\sqrt{r_lN}\gamma_a}.
\end{equation}
Since entropy of a pure state is zero, the Holevo quantity is just the entropy of the average ensemble's state
\begin{multline}
    \chi\left(\mathcal{E}\right)
    =
    S\left(\frac{1}{2}\hat{\rho}^{(0)}+\frac{1}{2}\hat{\rho}^{(1)}\right)
    \\=
    h_2\left(\frac{1}{2}-\frac{1}{2}\left|\braket{\sqrt{r_lN}\gamma_0\,|\, \sqrt{r_lN}\gamma_1}\right|\right)
    \\=
    h_2\left(\frac{1}{2}-\frac{1}{2}\exp\left[
    -\frac{1}{2}\cdot r_lN \left( \gamma_1 - \gamma_0 \right)^2
    \right]\right),
    \label{Holevo_No_PR}
\end{multline}
where $h_2(x)=x\cdot\log_2(x)-(1-x)\cdot\log_2(1-x)$ is binary entropy.
Applying phase randomization transforms the ensemble of pure coherent states into the mixtures of Fock states
\begin{equation}
\mathcal{E}_{\text{PR}}=\left\{\left(\frac{1}{2},\,\hat{\rho}^{(0)}_{\text{PR}}\right),\:\left(\frac{1}{2},\,\hat{\rho}^{(1)}_{\text{PR}}\right)\right\},
\end{equation}
\begin{multline}
    \hat{\rho}^{(a)}_{\text{PR}}
    =
    \frac{1}{2\pi}\int\limits_0^{2\pi}d\varphi
    \ket{\sqrt{r_lN}\gamma_a\cdot e^{i \varphi}}\!\bra{\sqrt{r_lN}\gamma_a \cdot e^{i \varphi}}
    \\= \sum\limits_{k=0}^{+\infty}p(k|a) \cdot \ket{k}\!\bra{k},
\end{multline}
where $p(k|a)$ is defined in Eq.\,(\ref{Poisson_probabilities}).
Now ensemble's states are diagonal, thus, quantum entropy is replaced with classical Shannon entropy
\begin{multline}
\chi\left(\mathcal{E}_{\text{PR}}\right)
    =
    S\left(\frac{1}{2}\hat{\rho}^{(0)}_{\text{PR}}+\frac{1}{2}\hat{\rho}^{(1)}_{\text{PR}}\right)
    -
    \frac{1}{2}S\left(\hat{\rho}^{(0)}_{\text{PR}}\right)
    -
    \frac{1}{2}S\left(\hat{\rho}^{(1)}_{\text{PR}}\right)\\
    = H\left( \bigg\{ \frac{p(k|0)\!+\!p(k|1)}{2} \bigg\}_{k=0}^{+\infty} \right)\!\\-\!\frac{1}{2} \! \left[ H \left( \bigg\{ p(k|0) \bigg\}_{k=0}^{+\infty} \right)\!+\!H \left( \bigg\{ p(k|1)\bigg\}_{k=0}^{+\infty} \right) \right].
    \label{Holevo_PR}
\end{multline}
Carrying out trivial mathematical transformations, one may conclude that the Holevo quantity for the phase randomization case Eq.\,(\ref{Holevo_PR}) coincides with the information obtained in ideal photon number measurement Eq.\,(\ref{information_photon_number}).
Figure\,\ref{dozens} also shows that the Holevo quantity for pure coherent states Eq.\,(\ref{Holevo_No_PR}) appeared to be much higher than considered photon number measurements, meaning that Eve may potentially utilize information about the phase to conduct more effective measurement.
It prompts legitimate users to implement phase randomization in their QKD scheme.

\begin{figure}[t]
\noindent\centering{
\includegraphics[width=0.8\columnwidth]{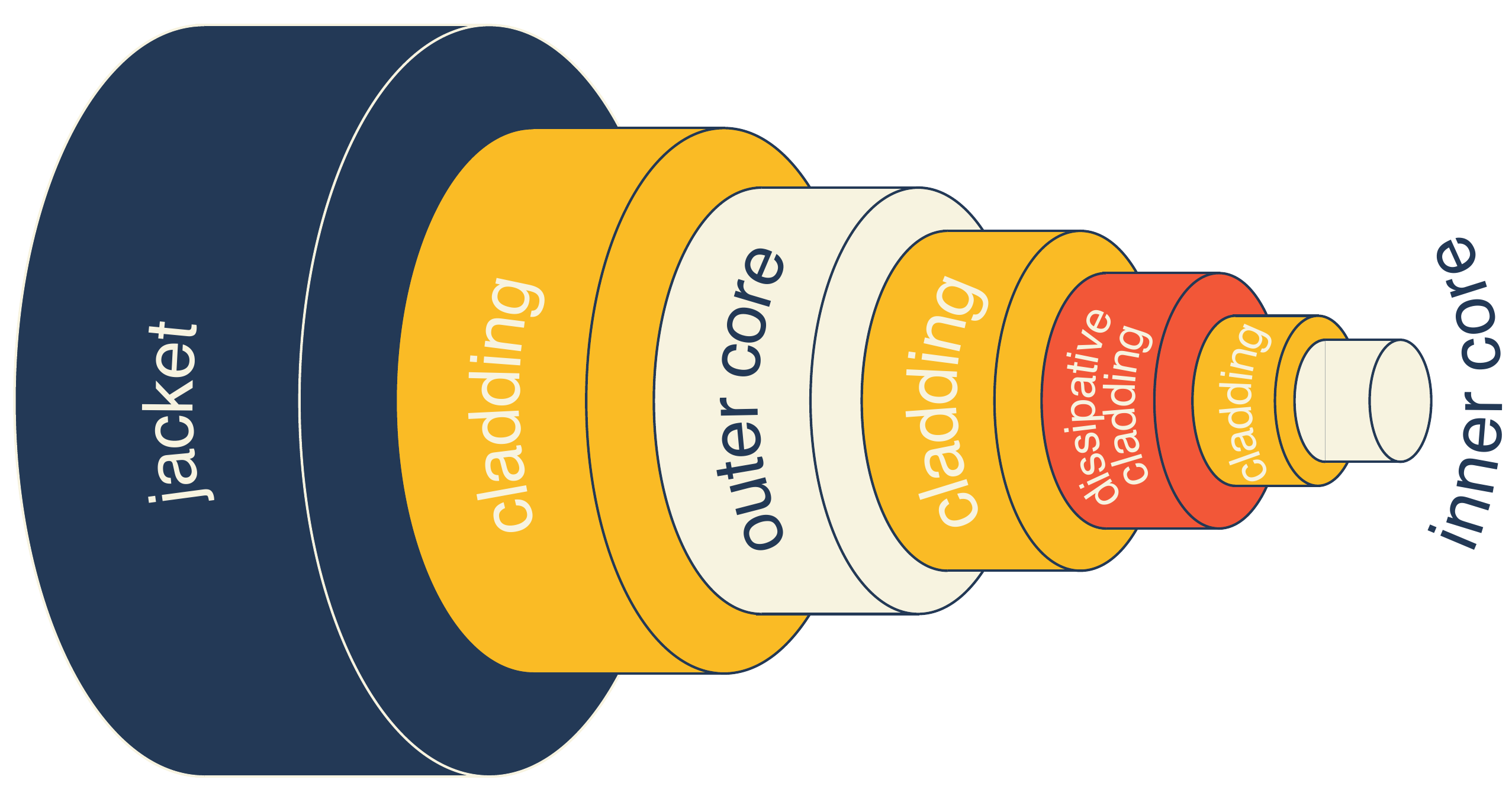}}
\caption{
\textbf{Cable design.} The cable namely includes (i) the inner core for information transmission and physical loss control, (ii) dissipative cladding converting scattering losses from the inner core into heat and completely destructing their information contents, and (iii) outer core for monitoring line integrity.
}
\label{Fig2_sup}
\end{figure}

\section*{NOTE 2. Controlled dissipation of natural losses and advanced line tomography}
In this note, we turn our attention to a particular cable design and advanced line tomography that transforms scattered photons into heat in a controlled manner. 
This method effectively precludes Eve from exploiting the natural losses. \\

The cable design transforming scattering losses into heat is sketched in Fig.\,\ref{Fig2_sup}. 
The inner fiber core carrying the information light pulses is surrounded by a cladding with a smaller refractive index and then by the dissipative cladding made out of metal or metal-doped silica. 
The dissipative cladding screens the scattering losses escaping the fiber core since the scattered wave undergoes the inelastic secondary scattering and transforms into heat. 
The resulting dissipative thermal losses cannot be deciphered even in principle. 
The dissipative cladding is, in turn, coated by the second standard refractive cladding layer, surrounded by an outer hollow fiber core and the outside refractive cladding.
The outer core enables legitimate users to perform constant reflectometry and transmittometry, thus exercising control over the dissipation of the scattering losses.
To remove the dissipative cladding for collecting the scattering losses or to create an artificial leakage from the inner core, Eve first needs to get through the outer core. 
The action, however, will not go unnoticed because of the permanent Alice and Bob's updating of the outer core's loss profile. 
Finally, the structure is surrounded by an outer jacket which may include a strengthening layer.
To mitigate local losses on fiber connections, they should be made spliced.

Physical loss control allows for a quantitative estimate and localization of any damage inflicted upon the outer or inner cores, see an experimentally obtained example of a reflectogram in Fig.\,4 of the main text, presenting the dependence of the backscattered power upon the distance to the scattering point: the linear regions represent homogeneous scattering along the line, while the sharp peaks or drops correspond to the local losses at fiber contacts and bends.
As opposed to the regular optical fiber, this specific line design eliminates the possibility of an undetected collection of scattering losses.
Any local intrusion brings Eve a negligible number of scattered photons; then, to collect a sufficient amount of the scattering losses, she needs to breach a large section of the controlled outer core, and before it is done, the legitimate users terminate the protocol.
Yet, our approach---particularly, the protocol outlined in the main text---resists a significant signal leakage fraction without the necessity of terminating transmission at first sight of the channel breach.
Instead, the legitimate users adapt the encoding and post-processing parameters based on the evaluation of the fraction of the signal possibly leaked to Eve, the outer and inner cores being monitored separately.
Importantly, although the inner core of the line is controlled only at step 1 of the protocol, legitimate users must control the outer core at all steps of the protocol.
The users can count scattering losses that leaked from the breached outer core region as stolen by Eve and then act accordingly, i.e., adapt or terminate the protocol.
In other words, the scattering losses escaping the breached outer core region can be equated to the artificial leakages directly from the inner core.
If the outer core is breached at the same spot where the inner core has an irredeemable local leakage, such as a bend, the users must take that this leakage is seized by Eve.

\section*{NOTE 3. Signal amplification}\label{AmplifiersNote}
In this note, we address optical amplification using the formalism of quantum channels.
We develop a mathematical representation of a sequence of optical amplifiers, which we later use for modeling a general beam splitter attack on the transmission.
Using this representation, we calculate the amplification-induced noise.

\subsection*{3.1 $\quad$  Amplification in doped fibers and losses}
In Er/Yt doped fiber, the photonic mode propagates through the inverted atomic medium.
To keep the medium inverted, a seed laser of a different frequency co-propagates with the signal photonic mode in the fiber and is then filtered out at the output by means of wavelength-division multiplexing (WDM).
The interaction between the inverted atoms and propagating light field mode $\hat{a}$ is given by the Hamiltonian  in the rotating wave approximation
\begin{equation}
    \hat{H}_{\text{int}} \!=\! i\kappa
    \sum\limits_{n=1}^{N}\!\left(
    \hat{a}^\dag  \hat{\sigma}^{(n)}_{-}
    -
    \hat{a}\hat{\sigma}^{(n)}_{+}
    \right)
    \!=\!
    i\kappa\!\left(
    \hat{a}^\dag \hat{S}_{-}
    -
    \hat{a}\hat{S}_{+}
    \right)
    ,
\end{equation}
\begin{equation}
    \hat{\sigma}^{(n)}_{-}
    =
    \ket{0}\bra{1}_{n}
    ,\,
    \hat{\sigma}^{(n)}_{+}
    =
    \ket{1}\bra{0}_{n}, \,
    \hat{S}_{\pm} = \sum \limits_{n=1}^N \hat{\sigma}^{(n)}_\pm,
\end{equation}
where we enumerate the atoms by index $n$, with $N$ being the overall number of atoms in the medium ($N\gg 1$), $\kappa$ is the interaction constant.
Here, each medium's atom is assumed to be a two-level system with its basis states $\ket{0}$ and $\ket{1}$ denoting the ground and excited states, respectively; $\hat{\sigma}^{(n)}_\pm$ is the $n$-th atom raising/lowering operator, while $\hat{S}_{\pm}$ is the collective raising/lowering operator, that shifts the number of excited atoms in the medium by one.
To simplify further calculations, we utilize the Holstein-Primakoff \cite{holstein} transformation which provides mapping between the collective ($\hat{S}_{+},\hat{S}_{-}$) and boson ($\hat{b},\hat{b}^\dag$) operators
\begin{equation}
    \hat{S}_+ = \sqrt{N} \sqrt{1-\frac{\hat{b}^\dag \hat{b}}{N}} \hat{b}, \quad \hat{S}_- = \sqrt{N} \hat{b}^\dag \sqrt{1-\frac{\hat{b}^\dag \hat{b}}{N}},
\end{equation}
where $\left[ \hat{b},\hat{b}^\dag \right]=1$.
In the regime, where average number of excitations is much smaller than the number of atoms $\langle \hat{b}^\dag \hat{b} \rangle \ll N$, one may use the approximation
\begin{equation}
    \hat{S}_{+} \approx \sqrt{N} \hat{b}, \quad
    \hat{S}_{-}  \approx \sqrt{N} \hat{b}^{\dagger}.
\end{equation}
The initial state of fully inverted atomic medium is $\ket{1}_{1} \!\otimes \ket{1}_2...\otimes \ket{1}_{N}$ (all $N$ atoms are in the excited state).
The Holstein-Primakoff transformation maps it to the vacuum  $\ket{0}_b$ (i.e. no excitations in the boson mode $b$).
The amplifier's medium with $m$ atoms in the ground state is now described by the state with $m$ excitations $\ket{m}_b$ that obeys the standard annihilation and creation operations
\begin{equation}
    \begin{gathered}
        \hat{b}\ket{m}_b = \sqrt{m}\ket{m-1}_b,\\
        \hat{b}^\dag \ket{m}_b = \sqrt{m+1}\ket{m+1}_b.
    \end{gathered}
\end{equation}
With that we have
\begin{equation}
    \hat{H}_{\text{int}}
    \approx
    i\kappa \sqrt{N}\left(\hat{a}^{\dagger}\hat{b}^{\dagger}
    -
    \hat{a}\hat{b}
    \right).
\end{equation}
The evolution operator of a propagating photon is given by
\begin{equation}
\label{U_g_sup}
    \hat{U}_g = e^{-i\hat{H}_{\text{int}} t/\hbar}=e^{g (\hat{a}^\dag \hat{b}^\dag - \hat{a} \, \hat{b})},
\end{equation}
where $g = \kappa\sqrt{N}t/\hbar$ and $t$ is effective time of interaction between the photonic mode and atomic medium.
Besides considering the channel acting on the propagating state, we also have to consider a conjugate channel acting on the creation operator $\hat{a}$ (it will be needed in the following cryptanalysis)
\begin{equation}
\label{eq: canonical amp}
\text{Amp}^*_{G}[\hat{a}]=\hat{U}_g^\dag\,  \hat{a} \, \hat{U}_g = \cosh(g) \hat{a} + \sinh(g) \hat{b}^\dag.
\end{equation}
In practice, the performance of erbium-doped fiber amplifiers (EDFAs) suffers from technical limitations, which arise in addition to the amplification limits on added quantum noise. 
These limitations are mainly caused by two factors: (i) the atomic population may be not completely inverted throughout the media, (ii) there may be coupling imperfection between the optical mode and the doped fiber section or main part of the fiber.
We will imply that these factors are accounted for in the loss channel prior to the amplification channel, as shown in Ref.\,[\onlinecite{Sanguinetti}].
The canonical transformation of the loss channel is
\begin{equation}
    \begin{gathered}
        \text{Loss}_T^*[\hat{a}] = \hat{\bar U}_T^\dag \hat{a} \hat{\bar U}_T  = \sqrt{T} \hat{a} + \sqrt{1-T} \hat{c},\\
        T = \cos^2 \lambda,
    \end{gathered}
\end{equation}
where $\lambda$ is the interaction parameter, $T$ is the proportion of the transmitted signal, the annihilation operator $\hat{c}$ corresponds to the initially empty mode which the lost photons go to, and $\hat{\bar U}_\lambda=e^{\lambda \hat{a}^\dag \hat{c} - \lambda \hat{a} \, \hat{c}^\dag}$.\\

\subsection*{3.2 $\quad$  $P$-function and its evolution under amplification}
Consider a single photonic mode with bosonic operators $\hat a$ and $\hat a^\dag$ acting in the Fock space.
To understand the effect of the amplification on the bosonic mode state, we will use the $P$-function formalism allowing to express any density operator as a quasi-mixture of coherent states:
\begin{equation}
\hat{\rho} = \int d^2 \alpha \, \text{P}(\alpha) \ket{\alpha}\bra{\alpha},
\label{P-representation}
\end{equation}
where $d^2 \alpha\equiv d\text{Re}(\alpha)\,d\text{Im}(\alpha)$ and the quasi-probability distribution $\text{P}(\alpha)$ is not necessarily positive.
For a given state with the density matrix $\hat\rho$ the $P$-function can be written as 
\begin{equation}
\text{P}(\alpha) =\tr   :\delta(\hat a-\alpha): \hat{\rho},
\end{equation}
where
\begin{equation}
:\delta(\hat a-\alpha): = \frac{1}{\pi^2}\int d^2\beta\,  e^{\alpha \beta^* - \alpha^* \beta} e^{\beta \hat a^\dag} e^{-\beta^* \hat a},
\end{equation}
see Ref.\,[\onlinecite{Vogel}] for details.
Amplification is described by a quantum channel given by
\begin{multline}
    \text{Amp}_{G=\cosh^2(g)}: \hat{\rho} \mapsto \text{Amp}_{G}[\hat{\rho}]\\=\tr_b \left[ \hat{U}_g\,   \hat{\rho} \otimes \ket{0} \bra{0}_b\, \hat{U}_g^\dag \right],
\end{multline}
where $\hat{U}_g$ is defined by Eq.\,(\ref{U_g_sup}),  $g$ is the interaction parameter characterizing the amplifier, $G = \cosh^2(g)$ is the factor by which the intensity of the input signal is amplified, and annihilation operator $\hat b$ corresponds to the auxiliary mode starting in the vacuum states. 
see e.g.\,[\onlinecite{Sekatski}].

To see how the $P$-function of a state transforms under the optical amplification, consider a simple situation where the input signal is in the pure coherent state $\ket{\gamma}\bra{\gamma}$ with the corresponding initial $P$-function $\text{P}_i(\alpha)= \delta(\alpha-\gamma)$ (delta-function acting on the complex plane).
After the amplification the $P$-function becomes
\begin{equation}
P(\alpha,\gamma,g) = \tr  :\delta(\hat a-\alpha): \text{Amp}_{G}\left[ \ket{\gamma} \bra{\gamma} \right].
\end{equation}
Bearing in mind the canonical transformation of the amplifier channel from Eq.\,(\ref{eq: canonical amp}), we get Eq.\,(3) from the main text:
given a pure coherent input state (with complex amplitude $\gamma$), the output state is a mixture of normally distributed coherent states; the mean complex amplitude is $\sqrt{G} \gamma$ and the standard deviation is $(G-1)/\sqrt{2}$:
\begin{equation}
P(\alpha, \gamma, G) = \frac{1}{\pi (G-1)} \exp\left(-\frac{|\alpha- \sqrt{G} \gamma|^2}{G-1}\right).
\label{$P$-function}
\end{equation}
\\

\begin{figure}[t]
    \noindent\centering{
    \includegraphics[width=0.85\columnwidth]{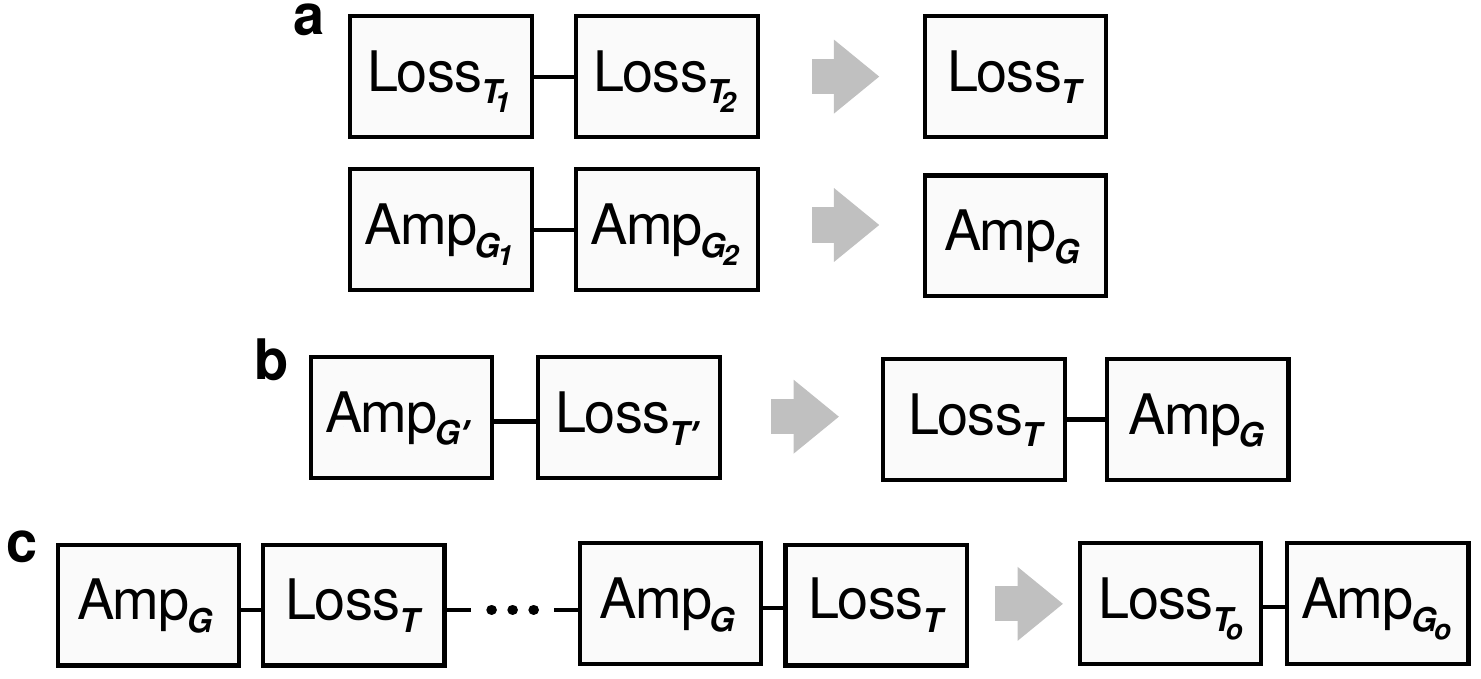}}
    \caption{
    \textbf{Compositions of loss and amplification channels and their equivalent representations.}
    \textbf{a} Two loss or amplification channels can be reduced to one. \textbf{b} Loss and amplification channels can be effectively rearranged. \textbf{c} A series of losses and amplifiers can be reduced to one pair of loss and amplification.
    } 
    \label{AmpLoss}
\end{figure}

\subsection*{3.3 $\quad$  Composition of amplifiers and losses}
In our quantum key distribution (QKD) protocol, the amplification is used to compensate the fiber losses. 
Long-distance transmission requires a cascade of amplifiers, in which case the signal’s evolution is determined by a sequence of multiple loss and amplification channels. 
In this section we prove that any such sequence can be mathematically reduced to a composition of one loss and one amplification channels. 
\\\\\textbf{Statement 1.} \textit{Two loss or amplification channels can be effectively reduced to one.} \smallskip\\
First, we show that a pair of loss or amplification channels can be effectively reduced to the one channel (Fig.\ref{AmpLoss}\textbf{a}).
To that end, let us consider two consequent loss channels:
\begin{multline}
    \left(\text{Loss}_{T_2}\circ \text{Loss}_{T_1}\right)^*\left[\hat{a}\right]
    \\=
    \sqrt{T_1T_2}\hat{a}+\sqrt{T_1(1-T_1)}\hat{c}_1+\sqrt{1-T_2}\hat{c}_2   
    \\
    =
    \sqrt{T_1T_2}\hat{a}+\sqrt{1-T_1T_2}\hat{c},
    \label{two_losses}
\end{multline}
where we defined operator $\hat{c}$
\begin{gather}
    \hat{c} =\frac{\sqrt{T_2(1-T_1)}\hat{c}_1 + \sqrt{1- T_2} \hat{c}_2}{\sqrt{1-T_1T_2}},
\end{gather}
acting on the vacuum state and satisfying the canonical commutation relation $[\hat{c},\hat{c}^\dag]=1$. 
The last expression of Eq.\,(\ref{two_losses}) represents the action of one loss channel with the effective parameter $T = T_1 T_2$
\begin{equation}
    \text{Loss}_{T_2}\circ \text{Loss}_{T_1}
    =
    \text{Loss}_{\left(T=T_1T_2\right)}.
\end{equation}
The same reasoning applies to amplifiers
\begin{equation}
    \text{Amp}_{G_2}\circ\text{Amp}_{G_1}=\text{Amp}_{G=G_1G_2}.
\end{equation}
\\\textbf{Statement 2.} \textit{Loss and amplification can always be represented as a composition where loss is followed by amplification}.\smallskip\\
Let us show that the composition of an amplification channel followed by a loss channel can be mathematically replaced with the pair of certain loss and amplification channels acting in the opposite order (Fig.\ref{AmpLoss}\textbf{b}).
Firstly, let us consider the transformation corresponding to the amplification followed by the loss 
\begin{multline}
    \left( \text{Loss}_{T'} \circ \text{Amp}_{G'} \right) ^*\left[\hat{a}\right]
    = \hat{U}_{g'}^\dag \hat{\bar U}_{\lambda'}^\dag \hat{a} \, \hat{\bar U}_\lambda \hat{U}_g \\
    =\sqrt{T' G'} \hat{a}+\sqrt{1-T'}\hat{c}+\sqrt{T'(1-G')} \hat{b}^\dagger.
\end{multline}
In the case of the opposite order we obtain
\begin{multline}
    (\text{Amp}_G \circ \text{Loss}_T)^*[\hat{a}] = \hat{\bar U}_\lambda^\dag \hat{U}_g^\dag \hat{a} \, \hat{U}_g \hat{\bar U}_\lambda=\\
    =\sqrt{T G} \hat{a} +\sqrt{G(1-T)} \hat{c} + \sqrt{G-1} \hat{b}^\dag.
\end{multline}
Considered transformations become identical when the parameters are related as
\begin{equation}
\begin{gathered}
    \text{Loss}_{T'}\circ \text{Amp}_{G'} = \text{Amp}_G \circ \text{Loss}_T,\\
    T = \frac{G' T'}{(G'-1)T'+ 1},\\
    G = (G'-1)T '+1.
    \label{permute}
\end{gathered}
\end{equation}
In other words, the two types of channels ”commute” provided that the parameters are modified in accord with these relations.
In particular, the parameters in the equation above are always physically meaningful $G \geq 1$, $0 \leq T \leq 1$, meaning that we can always represent loss and amplification in form of a composition where loss is followed by amplification (the converse is not true).
\\\\\textbf{Statement 3.} \textit{A series of losses and amplifiers can be effectively reduced to one pair of loss and amplification}.\smallskip\\
Let us finally show that the sequence of loss and amplification channels can be mathematically represented as one pair of loss and amplification  (Fig.\ref{AmpLoss}\textbf{c}).
Consider the transformation
\begin{equation}
    \Phi_M = (\text{Amp}_G\circ \text{Loss}_T)^{\circ M},
\end{equation}
corresponding to the series of $M$ identical loss and amplification channels, for which we want to find a simple representation. 
According to Statement 2, we can effectively move all losses to the right end of the composition, i.e., permute the channels in such a way that all the losses act before amplification.
Every time the loss channel with the transmission probability $T_{(i)}$ is moved before an amplifier with the amplification factor $G_{(i)}$, the parameters are transformed in accord with Eq.\,(\ref{permute}):
\begin{equation}
    \begin{gathered}
        T_{(i)} \mapsto T_{(i+1)}=\frac{G_{(i)} T_{(i)}}{(G_{(i)}-1) T_{(i)} + 1},\\
        G_{(i)} \mapsto G_{(i+1)}=(G_{(i)}-1)T_{(i)} +1.
    \end{gathered}
\end{equation}
In our sequence we can pairwise transpose all neighboring losses with amplifier (starting with the first amplifier and the second loss).
After repeating this operation $M-1$ times, bearing in mind the Statement\,1, we find that
\begin{multline}
        \Phi_M = \text{Amp}_{G_{(0)}}\circ \text{Amp}_{G_{(1)}}\circ \dots \circ\text{Amp}_{G_{(M-1)}}\\
        \quad \circ \text{Loss}_{T_{(M-1)}}\circ \text{Loss}_{T_{(M-2)}}\circ \dots \circ \text{Loss}_{T_{(0)}}
         \\= \text{Amp}_{G_\circ} \circ \text{Loss}_{T_\circ},
\end{multline}
where 
\begin{equation}
    T_\circ = \prod_{i=0}^{M-1} T_{(i)}, \quad
    G_\circ = \prod_{i=0}^{M-1} G_{(i)},
\end{equation}
i.e., the series of losses and amplifiers is equivalent to the loss channel of transmission $T_\circ$ followed by the amplifier with amplification factor $G_\circ$.
Note now that the value $\eta\equiv G_{(i)}T_{(i)}=G T$ cannot be changed by permutations. 
Let us define
\begin{equation}
    F_{(i)} = (G_{(i)}-1) T_{(i)}+1,
\end{equation}
and bear in mind that 
\begin{multline}
    F_{(i+1)} = (G_{(i+1)}-1) T_{(i+1)}+1
    \\= \frac{(F_{(i)}-1)}{F_{(i)}} TG + 1
    =  \eta \left(\frac{F_{(i)}-1}{F_{(i)}}\right) + 1.
\end{multline}
We can write
\begin{equation}
T_{(i+1)} = \frac{ T G}{F_{(i)}},\quad
G_{(i+1)} = F_{(i)},
\end{equation}
and
\begin{equation}
    \begin{gathered}
        G_\circ =  G \prod_{i=0}^{M-2}F_{(i)}, \quad
    T_\circ =  \frac{T (TG)^{M-1}}{ \prod_{i=0}^{M-2} F_{(i)}}= \frac{(T G)^M}{G_\circ}.
    \end{gathered}
\end{equation}
Let us find the explicit form of $G_\circ$ and $T_\circ$ by solving the recurrence relation.
Define $A_{n}$ and $B_{n}$ through the relation
\begin{equation}
    F_{(n-1)} = \frac{A_{n}}{B_{n}}.
    \label{def Fn}
\end{equation}
Then
\begin{equation}
    F_{(n+1)} = \frac{(\eta\!+\!1) F_{(n)}-\eta}{F_{(n)}} \\= \frac{(\eta\!+\!1) A_{n+1}-\eta B_{n+1} }{A_{n+1}}.
    \label{relation for Fn through An and Bn}
\end{equation}
It follows from Eqs.\,\eqref{def Fn} and \eqref{relation for Fn through An and Bn}  that $B_{n+1}= A_n $ and
\begin{equation}
    A_{n+1} = (\eta+1) A_n-\eta B_n = (\eta+1) A_n-\eta A_{n-1}.
\end{equation}
We see that the solution of this equation has a form
\begin{equation}
A_{n} = c_1 + c_2 \eta^n,
\end{equation}
where $c_1$ and $c_2$ are the constants, which are determined by $F_0 =(G-1)T+1$: we take $A_1 = (G-1)T+1$ and $A_0 =1$, and obtain
\begin{equation}
    c_1 =\frac{T-1}{G T-1}, \quad c_2 = \frac{(G-1) T}{G T-1}.
\end{equation}
Notably, the product $\Pi_{n=0}^{M-2} F_{(n)}$ appearing in the final expression becomes relatively simple
\begin{equation}
    \prod_{n=0}^{M-2} F_{(n)} = \frac{(G-1) (G T)^M+G (T-1)}{G (G T-1)},
\end{equation}
and we have
\begin{equation}
    \begin{gathered}
        \Phi_M = (\text{Amp}_G\circ \text{Loss}_T)^{\circ M}= \text{Amp}_{G_\circ} \circ \text{Loss}_{T_\circ},\\
    G_\circ = \frac{(G-1) (G T)^M+G (T-1)}{G T-1},\\
    T_\circ = \frac{(TG)^M}{G_\circ}.
    \end{gathered}
\end{equation}
The case of $TG=1$ is particularly interesting as the average photon number of the transmitted signal remains preserved (which is different from the total output photon number as it has the noise contribution). 
In the limit $G\to 1/T$ we have
\begin{equation}
    \begin{gathered}
        G_\circ = G(M(1-T)+T),\\
        T_\circ =\frac{T}{M (1-T)+T}.
    \end{gathered}
    \label{equivGT}
\end{equation}

\subsection*{3.4 $\quad$  Effective model of the line}
\begin{figure}[t]
    \noindent\centering{
    \includegraphics[width=0.95\columnwidth]{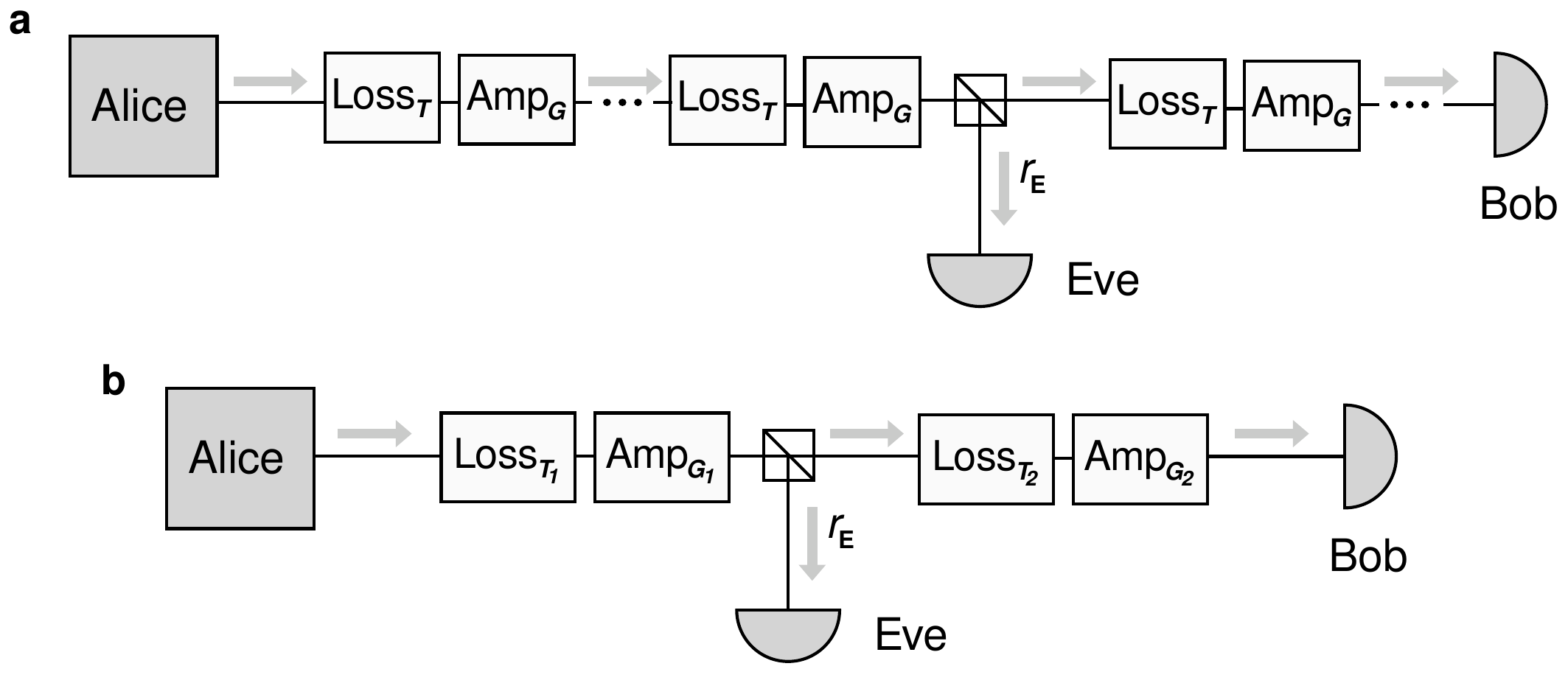}}
    \caption{
    \textbf{Schematic representation of a beam splitter attack.}
    \textbf{a} 
    Alice and Bob are connected by a quantum channel comprised of the composition of amplifiers and losses.
    Eve conducts a beam splitter attack seizing a part of the signal somewhere along the optical line.
    Thus, she divides the line into two parts.
    \textbf{b} An equivalent scheme: the losses and amplifiers before and after the point of Eve's intervention are represented by two pairs of loss and amplification channels defined by the parameters $\{T_1,\,G_1\}$ and $\{T_2,\,G_2\}$ respectively.
    } 
    \label{G1G2}
\end{figure}

We consider how Eve performs the beam splitter attack seizing the part of the signal somewhere along the optical line as shown in Fig.\ref{G1G2}\textbf{a}.
If the signal intensity incident to the beam splitter is 1, then intensity $r_{\text{E}}$ goes to Eve, and $1-r_{\text{E}}$ goes to Bob’s direction.
The proportion of the transmitted signal on the distance $d$ between two neighbouring amplifiers is determined by
\begin{equation}
    T = 10^{-\xi \cdot d}, \,\,\, G=\frac{1}{T},
\end{equation}
where $\xi=0.02\,\text{km}^{-1}$ is the parameter of losses typical for the optical fibers and $G$ is amplification factor of each amplifier.
Let $D_{\text{AB(AE)}}$ be the distance between Alice and Bob (Alice and Eve), then the numbers of amplifiers before and after the beam splitter $M_1$ and $M_2$ are given by
\begin{equation}
\begin{gathered}
    M_1 = D_{\text{AE}}/d,\\
    M_2 = (D_{\text{AB}}-D_{\text{AE}})/d.
\end{gathered}
\end{equation}
According to Statement 3, the scheme can be simplified by reducing the losses and amplifications before and after the beam splitter to two loss and amplification pairs with the parameters $\{T_1=1/G_1,\,G_1\}$ and $\{T_2=1/G_2,\,G_2\}$ respectively (Fig. \ref{G1G2}\textbf{b})
\begin{equation}
    G_1 = G \left( M_1 (1-T)+T \right) = \left( 10^{-\xi d}-1 \right) \cdot \frac{D_{\text{AE}}}{d}+1,
    \label{G1}
\end{equation}
\begin{multline}
    G_2 = G \left( M_2 (1-T)+T \right) \\= \left( 10^{-\xi d}-1 \right) \cdot \frac{D_{\text{AB}}-D_{\text{AE}}}{d}+1.
    \label{G2}
\end{multline}

\subsection*{3.5 $\quad$  Fluctuations}
Let us calculate the fluctuation of the number of photons in a pulse after it passes through a sequence of $M$ loss regions and amplifiers.
Let $|\gamma|^2$ be the input average photon number;
as follows from Eqs.\,\eqref{P-representation} and\,\eqref{$P$-function}, the average number of photons $n$ in the output signal is
\begin{equation}
\begin{split}
    n = \langle\hat{a}^\dagger\hat{a}\rangle = |\gamma|^2 + G_\circ - 1,
    \label{average_n}
\end{split}
\end{equation}
where $G_\circ$ is given by Eq.\,\eqref{equivGT}. 
The variance of the output photon number is
\begin{multline}
    \delta n = \sqrt{\langle(\hat{a}^\dagger\hat{a})^2\rangle - (\langle\hat{a}^\dagger\hat{a}\rangle)^2} \\    
    =\Big( M(G-1)(M(G-1)+1)\\+|\gamma|^2(2M(G-1)+1) \Big)^{1/2}
    \label{fluctuation of intensity after M amp-s}
\end{multline}
Note that even if $|\gamma|^2=0$, $n$ and $\delta n$ are still non-zero.
This particularly means that on top of the mode of interest, amplification also generates noise in other modes.
Assume that Bob has an optical filter with bandwidth $\Delta \nu$ on which the amplification factor is constant, and the detection time is $\tau\gg1/\Delta \nu$.
Then, the average number of photons due to noise from the secondary modes is 
\begin{equation}
    n_\text{noise}\simeq 2[G(M(1-T)+T)-1]\tau\Delta\nu,
\end{equation}
where factor 2 is due to two possible polarizations.
We will thoroughly address the 
issue of noise at different optical filtration regimes in our forthcoming experimental publication.

In the limit $|\gamma|^2\gg GM\gg1$ for the ideal optical filter transmitting only the signal mode we get
\begin{equation}
\label{fluct}
    \delta n\simeq \sqrt{n G M}.
\end{equation}
Provided that there are no other sources of noise, this quantity determines the precision of the physical loss control: if the test pulse carries $n$ photons, the minimum detectable leakage is
\begin{equation}
r_\text{E}^\text{min}\sim\sqrt{MGn}/n=\sqrt{MG/n}.
\end{equation}
The result coincides with the estimate given in Methods.

\section*{NOTE 4. Photon number encoding}
This note is devoted to the detailed description of the photon number encoding scheme in the context of the beam splitter attack.
We derive the respective amounts of information that the users and Eve know about the transmitted signal.
We also address correlations due to the optical amplification between Eve's and Bob's quantum states.\\

\subsection*{4.1 $\quad$  Phase randomization}
In the photon number encoding, we assume the phase of each signal pulse being completely random\,\cite{PhR_exp_zhao}, and unknown both to the eavesdropper and the users.
The phase alteration between consecutive signal pulses may be achieved through rebooting the light source after each pulse, or with an additional randomly acting phase modulator.
We take that Bob's performs solely the energy measurement of the incoming states without measuring the phases, so Alice does not send the phase reference.
In this case, Eve cannot measure the phases either, and the combined system of Alice's random bit (A), the signal incident to Bob (S) and Eve's seized state (E) can be expressed as a mixture of states averaged over all possible phases, 
\begin{multline}
    \left\langle\hat{\rho}_{\text{ASE}}^{\rightarrow\text{Bob}}\right\rangle_\varphi
    =
    \frac{1}{2\pi}\int\limits_{0}^{2\pi}\!\!d\varphi\,\,
    \cdot
    \frac{1}{2} \sum \limits_{a=0,1} \ket{a} \bra{a}_\text{A} \\\otimes \int\!\!d^2 \alpha\,\, 
    P(\alpha,\sqrt{T_1} e^{i\varphi}\gamma_a,G_1) 
    \\\times\int\!\!d^2\beta\,\,
    P\left( \beta,\sqrt{(1-r_{\text{E}})T_2} \alpha, G_2 \right) \\\times \ket{\beta} \bra{\beta}_\text{S}  \otimes \ket{\sqrt{r_\text{E}} \alpha} \bra{\sqrt{r_\text{E}} \alpha}_\text{E}
    .
    \label{app_ic_ase}
\end{multline}
It follows from Eq.\,(\ref{$P$-function}) that for any $\varphi\in\mathbb{R}$ we have $P\left(x,ye^{i\varphi},z\right)=P\left(xe^{-i\varphi},y,z\right)$.
We thus can write
\begin{multline}
    \left\langle\hat{\rho}_{\text{ASE}}^{\rightarrow\text{Bob}}\right\rangle_\varphi
    =
    \frac{1}{2} \sum \limits_{a=0,1} \ket{a} \bra{a}_\text{A} \otimes 
    \int\!\!d^2\alpha\,\, P(\alpha,\sqrt{T_1}\gamma_a,G_1)
    \\\times
    \int\!\!d^2\beta\,
    P\left( \beta,\sqrt{(1\!-\!r_{\text{E}})T_2} \alpha, G_2 \right)
    \cdot
    \frac{1}{2\pi} 
    \int\limits_{0}^{2\pi}\!\!d\varphi\,
    \ket{e^{i\varphi}\beta}\! \bra{e^{i\varphi}\beta}_\text{S}
    \\\otimes \ket{\sqrt{r_\text{E}} e^{i\varphi}\alpha} \bra{\sqrt{r_\text{E}} e^{i\varphi}\alpha}_\text{E}.
    \label{ASE_final}
\end{multline}

\subsection*{4.2 $\quad$  Bob's information}
We first estimate the mutual information between Alice and Bob, which is $S(\text{A})-S(\text{A}|\text{B})$ in Eq.\,(6) of the main text.
For that purpose, we trace out Eve's subsystem and consider a bipartite quantum state shared between legitimate users right before Bob's measurement is conducted $\left\langle\hat{\rho}^{\rightarrow \text{Bob}}_\text{AS}\right\rangle_{\varphi}=\tr_{\text{E}}\left\langle\hat{\rho}_{\text{ASE}}^{\rightarrow\text{Bob}}\right\rangle_\varphi$
\begin{multline}
    \left\langle
    \hat{\rho}^{\rightarrow \text{Bob}}_\text{AS}\right\rangle_{\varphi} = \frac{1}{2} 
    \sum \limits_{a=0,1} \ket{a} \bra{a}_\text{A}\otimes 
    \int\!\!d^2\alpha\,\,
    P(\alpha,\sqrt{T_1} \gamma_a,G_1)  
    \\
    \times \!\! \int\!\!d^2\beta P \!\left( \beta,\sqrt{(1-r_{\text{E}})T_2} \alpha, G_2 \right) \\
   \times \frac{1}{2\pi} \!\!
    \int \limits_0^{2 \pi}\!\!d\varphi \ket{\beta e^{i \varphi}} \bra{\beta e^{i \varphi}}_\text{S}.
\end{multline}
Given that Alice sent bit $a \in \{ 0,1 \}$, the probability that Bob's measurement outcome is $b \in \{0,1,\mbox{fail} \}$ can be written as
\begin{multline}
    p(b|a) = \int\frac{d^2 \alpha}{\pi (G_1 -1 )} 
    \cdot \exp \left( -\frac{\left| \alpha - \sqrt{G_1 T_1} \gamma_a \right|^2}{G_1 -1 } \right)
    \\\times\int\frac{d^2 \beta}{\pi (G_2 - 1)} 
    \,\exp \left( -\frac{\left| \beta - \sqrt{(1-r_{\text{E}})G_2 T_2} \alpha \right|^2}{G_2 -1 } \right) \\\times \frac{1}{2\pi}
    \int\limits_0^{2\pi}\!\!d\varphi \bra{\beta e^{i \varphi}} \hat{E}_b \ket{\beta e^{i \varphi}}.
    \label{cond_first}
\end{multline}
The probability of finding $k$ photons in the coherent state $\ket{\beta e^{i \varphi}}$ is defined by the Poisson distribution and depends only on $|\beta|$
\begin{equation}
    \left| \braket{k|\beta e^{i \varphi}} \right|^2 = \frac{|\beta|^{2k}e^{-|\beta|^2}}{k!} = \left| \braket{k|\beta } \right|^2,
\end{equation}
which particularly means that for any $b \, \bra{\beta e^{i \varphi}} \hat{E}_b \ket{\beta e^{i \varphi}} = \bra{\beta} \hat{E}_b \ket{\beta}$.
Thus, the probability of results ``0'' and ``1'' can be expressed  analytically
\begin{multline}
    \bra{\beta e^{i \varphi}} \hat{E}_1 \ket{\beta e^{i \varphi}} = \bra{\beta} \hat{E}_1 \ket{\beta} = \sum \limits_{k=\mu+\theta_2}^{\mu+\theta_4} \frac{|\beta|^{2k} e^{-|\beta|^2}}{k!}\\
    =\frac{\Gamma\left( \mu+\theta_4+1,|\beta|^2 \right)}{\Gamma\left( \mu+\theta_4+1 \right)} - \frac{\Gamma\left( \mu+\theta_2,|\beta|^2 \right)}{\Gamma\left( \mu+\theta_2 \right)},
\end{multline}
\begin{multline}
    \bra{\beta} \hat{E}_0 \ket{\beta} =
    \bra{\beta} \hat{E}_0 \ket{\beta} = \sum \limits_{k=\mu-\theta_3}^{\mu-\theta_1} \frac{|\beta|^{2k} e^{-|\beta|^2}}{k!}\\
    \frac{\Gamma\left( \mu-\theta_1+1,|\beta|^2 \right)}{\Gamma\left( \mu-\theta_1+1 \right)}-\frac{\Gamma\left( \mu-\theta_3,|\beta|^2 \right)}{\Gamma\left( \mu-\theta_3 \right)},
\end{multline}
where $\Gamma(z)$ is the Euler gamma function and $\Gamma(z_1,z_2)$ is the incomplete gamma function.
Turning again to Eq.\,(\ref{cond_first}), we can change the integration order using Fubini's theorem. 
The analytical integration over $d^2 \alpha$ gives us
\begin{multline}
    p(b|a) = \frac{e^{-\frac{\gamma_a^2(1-r_{\text{E}})}{G_1+G_2-2-r_{\text{E}}(G_1-1)}}}{\pi \left( G_1+G_2-2-r_{\text{E}}(G_1-1) \right)} 
    \\\times\int\limits_0^{+\infty}\!\!d|\beta|\,\,|\beta|\cdot e^{-\frac{|\beta|^2}{G_1+G_2-2-r_{\text{E}}(G_1-1)}} 
    \bra{\beta} \hat{E}_b \ket{\beta} 
    \\\times\int\limits_0^{2\pi}\!\!d\varphi_{\beta}\,\, \exp \left(  \frac{2\sqrt{1-r_{\text{E}}}\gamma_a |\beta| \cos \varphi_\beta}{G_1+G_2-2-r_{\text{E}}(G_1-1)} \right),
\end{multline}
where $\varphi_\beta=\text{arg}(\beta)$.
The rightmost integral can be reduced to the modified Bessel function of the first kind:
\begin{multline}
\int\limits_0^{2\pi}\!\!d\varphi_{\beta}\,\exp\left({  \frac{2\sqrt{1-r_{\text{E}}}\gamma_a |\beta| \cos \varphi_\beta}{G_1+G_2-2-r_{\text{E}}(G_1-1)}}\right) \\= 2\pi \cdot I_0\left( \frac{2\sqrt{1-r_{\text{E}}}\gamma_a |\beta|}{G_1+G_2-2-r_{\text{E}}(G_1-1)} \right).
\end{multline}
Thus, we have
\begin{multline}
    p(b|a) = \frac{2e^{-\frac{\gamma_a^2(1-r_{\text{E}})}{G_1+G_2-2-r_{\text{E}}(G_1-1)}}}{G_1+G_2-2-r_{\text{E}}(G_1-1)}
    \\\times\int\limits_{0}^{+\infty}\!\!d|\beta|\,\,|\beta| \cdot \bra{\beta} \hat{E}_b \ket{\beta} \cdot e^{-\frac{|\beta|^2}{G_1+G_2-2-r_{\text{E}}(G_1-1)}} 
    \\\times 
    I_0\left( \frac{2\sqrt{1-r_{\text{E}}}\gamma_a |\beta|}{G_1+G_2-2-r_{\text{E}}(G_1-1)} \right).
\end{multline}
The mutual information between Alice and Bob after the post-selection can be calculated as
\begin{multline}
    I\left(\text{A,\,B}\right)
    \equiv
    S\left(\text{A}\right)
    -
    S\left(\text{A}|\text{B}\right)
    =
    S\left(\text{A}\right)
    +
    S\left(\text{B}\right)
    -
    S\left(\text{AB}\right)
    \\
    \!\!\!\!\!=
    h_2\left(
    \sum\limits_{b=0,1}
    \!\!\frac{p(b|0)}{2p_{\checkmark}}
    \!\right)
    +
    h_2\left(
    \sum\limits_{a=0,1}
    \!\!\frac{p(0|a)}{2p_{\checkmark}}
    \!\right)
    \\+
    \sum\limits_{a=0,1}
    \sum\limits_{b=0,1}
    \frac{p(b|a)}{2p_{\checkmark}}
    \log_2\left(\frac{p(b|a)}{2p_{\checkmark}}\right),\!
\end{multline}
where $h_2(p)=-p\cdot \log_2 p -(1-p)\cdot \log_2(1-p)$ is the binary entropy and the probability of the successful measurement outcome is
\begin{equation}
    p_{\checkmark}
    =
    \frac{1}{2}\sum\limits_{a=0,1}
    \sum\limits_{b=0,1}p(b|a).
\end{equation}
\\

\subsection*{4.3 $\quad$  Eve's information}
To estimate Eve's information ($I(\text{A:E})$ in the Eq.\,(6) of the main text), we find the explicit form of the quantum state owned by Eve after the users perform post-selection.
The density matrix of the joint ABE system is
\begin{multline}
    \left\langle\hat{\rho}_{\text{ABE}}^{\,f}\right\rangle_\varphi
    =
    \sum \limits_{b=0,1}
    \sum\limits_{a=0,1}
    \frac{1}{2p(\checkmark|a)}
    \int\!\!d^2\alpha\,\, P(\alpha,\sqrt{T_1}\gamma_a,G_1)
    \\\times
    \ket{a} \bra{a}_\text{A}
    \otimes
    \ket{b}\bra{b}_{\text{B}}\otimes
    \frac{1}{2\pi} \int\limits_{0}^{2\pi}\!\!d\varphi\,\,\left|e^{i\varphi}\sqrt{r_{\text{E}}}|\alpha|\right\rangle \left\langle e^{i\varphi} \sqrt{r_{\text{E}}}|\alpha| \right| _\text{E}
    \\\times
    \int\!\!d^2\beta\,\,P\left( \beta,\sqrt{(1-r_{\text{E}})T_2} \alpha, G_2 \right)
    \braket{\beta|\hat{E}_b|\beta}
    .
\end{multline}
For the further calculations it is useful to introduce ``conditional'' Eve's density matrix, i.e., Eve's density matrix in the case that Alice sent bit $a$ and Bob got a successful measurement outcome:
\begin{equation}
    \hat{\rho}_{\text{E}}^{(a)}
    \equiv
    \text{tr}_{\text{AB}}
    \left[
    \left(2 \ket{a} \bra{a}_{\text{A}}\otimes\hat{\mathds{1}}\otimes\hat{\mathds{1}}\right)
    \cdot
    \left\langle\hat{\rho}_{\text{ABE}}^{\,f}\right\rangle_{\varphi}
    \right].
\end{equation}
The explicit form of this matrix is
\begin{multline}
    \hat{\rho}_{\text{E}}^{(a)}
    =
    \frac{1}{p(\checkmark|a)}
    \int\!\!d^2\alpha\,\,
    P\left(\alpha;\sqrt{T_1}\gamma_a,G_1\right)
    \\\times
    \int\!\!d^2\beta\,\,
    P\left(\beta;
    \sqrt{(1-r_{\text{E}})T_2}\alpha,G_2\right)
    \braket{\beta|\hat{E}_{\checkmark}|\beta}
    \\\times
    \left[
    \frac{1}{2\pi}\int\limits_{0}^{2\pi}\!\!d\varphi\,\,
    \left|e^{i\varphi}\sqrt{r_{\text{E}}}|\alpha|\right\rangle \left\langle e^{i\varphi} \sqrt{r_{\text{E}}}|\alpha| \right|_\text{E}\right],
    \label{eve_matr_5ints}
\end{multline}
where $\hat{E}_{\checkmark}\equiv\hat{E}_0+\hat{E}_1$.
To simplify the expression, one can convert integration into Polar coordinates ($\alpha=|\alpha|e^{i\varphi_{\alpha}}$, $d^2\alpha=|\alpha|\,d|\alpha|\,d\varphi_{\alpha}$), apply Fubini's theorem, and carry out integration over $\varphi_{\alpha}$ analytically
\begin{multline}
    \int\limits_{0}^{2\pi}\!\!d\varphi_{\alpha}\,\,
    P\left(|\alpha|e^{i\varphi_{\alpha}};\sqrt{T_1}\gamma_a,G_1\right)
    \\=
    \frac{2\exp\left(-\frac{|\alpha|^2+|\gamma_a|^2}{G_1-1}\right)}{G_1-1}\cdot I_0\left(\frac{2|\alpha|\gamma_a}{G_1-1}\right),
    \label{int1}
\end{multline}
where $I_0(z)$ is the modified Bessel function of the first kind.
The same procedure can be carried out for integration over $\beta$: bearing in mind that $\braket{\beta|\hat{E}_{\checkmark}|\beta}=\braket{\beta|\hat{E}_{1}|\beta}+\braket{\beta|\hat{E}_{0}|\beta} \equiv f(|\beta|)$ does not depend on $\varphi_{\beta}$, we have
\begin{multline}
    \int\limits_{0}^{2\pi}\!\!d\varphi_{\beta}\,\,
    P\left(|\beta|e^{i\varphi_{\beta}};\sqrt{(1-r_{\text{E}})T_2}\alpha,G_2\right)
    \\
    \!\!=
    \frac{2\exp\left(-\frac{|\beta|^2+(1-r_{\text{E}})|\alpha|^2}{G_2-1}\right)}{G_2-1} I_0\!\left(\frac{2|\beta|\sqrt{1-r_{\text{E}}}|\alpha|}{G_2-1}\right)\!.
    \label{int2}
\end{multline}
The rightmost integral in Eq.\,(\ref{eve_matr_5ints}) also can be calculated in the analytical way and expressed in terms of the Fock states $\left\{\ket{n}\right\}_{n=0}^{+\infty}$:
\begin{multline}
    \frac{1}{2\pi}\int\limits_{0}^{2\pi}\!\!d\varphi\,\,
    \left|e^{i\varphi}\sqrt{r_{\text{E}}}|\alpha|\right\rangle \left\langle e^{i\varphi} \sqrt{r_{\text{E}}}|\alpha| \right|
    \\=
    e^{-r_{\text{E}}|\alpha|^2}\sum\limits_{n=0}^{+\infty}\frac{\left(r_{\text{E}}|\alpha|^2\right)^n}{n!}\ket{n}\bra{n}.
    \label{int3}
\end{multline}
Substituting the results from Eqs.\,(\ref{int1}--\ref{int3}) into Eq.\,(\ref{eve_matr_5ints}), we get
\begin{multline}
    \hat{\rho}_{\text{E}}^{(a)}
    =
    \frac{4/p(\checkmark|a)}{(G_1-1)(G_2-1)}
    \int\limits_{0}^{+\infty}\!\!d|\alpha|\,\,|\alpha|
    \exp\left(-\frac{|\alpha|^2+|\gamma_a|^2}{G_1-1}\right)
    \\\times I_0\left(\frac{2|\alpha|\gamma_a}{G_1-1}\right)
    \sum\limits_{n=0}^{+\infty}
    \exp\left(-r_{\text{E}}|\alpha|^2\right)\frac{\left(r_{\text{E}}|\alpha|^2\right)^n}{n!}\ket{n}\bra{n}
    \\\times
    \int\limits_{0}^{+\infty}\!\!d|\beta|\,\,|\beta|
    \exp\left(-\frac{|\beta|^2+(1-r_{\text{E}})|\alpha|^2}{G_2-1}\right)
    \\\times I_0\left(\frac{2|\beta|\sqrt{1-r_{\text{E}}}|\alpha|}{G_2-1}\right)
    f(|\beta|).
\end{multline}
Note that the resulting density matrix is diagonal in the Fock basis---which is natural given the phase randomization. 
The diagonal elements of the matrix can be expressed as
\begin{multline}
    \braket{n|\hat{\rho}_{\text{E}}^{(a)}|n}
    =
    \frac{4r_{\text{E}}^n\exp\left(-\frac{\gamma_a^2}{G_1-1}\right)}{n!(G_1-1)(G_2-1)p(\checkmark|a)}
    \\\times\int\limits_{0}^{+\infty}\!\!d|\beta|\,\,|\beta|\cdot f(|\beta|)\exp\left(-\frac{|\beta|^2}{G_2-1}\right)
    \\\times\int\limits_{0}^{+\infty}\!\!d|\alpha|\,\,|\alpha|^{2n+1}
    \exp\left(-|\alpha|^2\cdot\left[\frac{1}{G_1-1}+\frac{1-r_{\text{E}}}{G_2-1}+r_{\text{E}}\right]\,\right)
    \\\times
    I_0\left(\frac{2|\alpha|\gamma_a}{G_1-1}\right)\cdot
    I_0\left(\frac{2|\beta|\sqrt{1-r_{\text{E}}}|\alpha|}{G_2-1}\right).
\end{multline}
In order to take integral over $|\alpha|$ analytically, we utilize the fact that the main contribution to the integral comes from $|\alpha|\gg 1$ which enables us to use the asymptotic expansion\,\cite{Kumer}:
\begin{equation}
    I_0(z)
    =
    \frac{e^z}{\sqrt{2\pi z}}\left(1+\frac{1}{8z}+O\left(\frac{1}{|z|^2}\right)\right),\; z\in\mathbb{R}.
    \label{approx_bessel}
\end{equation}
Thus, we have
\begin{multline}
    I_0\left(\frac{2|\alpha|\gamma_a}{G_1-1}\right)\cdot
    I_0\left(\frac{2|\beta|\sqrt{1-r_{\text{E}}}|\alpha|}{G_2-1}\right)=
    \frac{1}{4\pi|\alpha|}
    \\
    \sqrt{\frac{(G_1-1)(G_2-1)}{\gamma_a|\beta|\sqrt{1-r_{\text{E}}}}} \exp\left(2|\alpha|\cdot\left(\frac{\gamma_a}{G_1-1}+\frac{|\beta|\sqrt{1-r_{\text{E}}}}{G_2-1}\right)\right)
    \\
    \!\!\!\times \left(1\!+\!\frac{1}{16|\alpha|}\!\left[\frac{G_1-1}{\gamma_a}\!+\!\frac{G_2-1}{|\beta|\sqrt{1-r_{\text{E}}}}\right]\!+\!O\left(\!\frac{1}{|\alpha|^2}\!\right)\!\right)\!.
\end{multline}
Utilizing this form, we get
\begin{multline}
    \braket{n|\hat{\rho}_{\text{E}}^{(a)}|n}
    \approx
    \frac{r_{\text{E}}^n\exp\left(-\frac{\gamma_a^2}{G_1-1}\right)/p(\checkmark|a)}{\pi\sqrt{\gamma_a|\beta|\sqrt{1-r_{\text{E}}}(G_1-1)(G_2-1)}}
    \\\times\int\limits_{0}^{+\infty}\!\!d|\beta|\,\,|\beta|\cdot f(|\beta|)\cdot e^{-\frac{|\beta|^2}{G_2-1}}
    \\
    \!\!\times \!\left(\varkappa_n\left(|\beta|\right)\!+\!\frac{\tilde{\varkappa}_n\left(|\beta|\right)}{16}\left[\frac{G_1\!-\!1}{\gamma_a}+\frac{G_2\!-\!1}{|\beta|\sqrt{1\!-\!r_{\text{E}}}}\right]\right)
    ,
    \label{rhodiag}
\end{multline}
where we introduced two subsidiary functions $\varkappa_n\left(|\beta|\right)$ and $\tilde{\varkappa}_n\left(|\beta|\right)$, $n\in\mathbb{N}$:
we define
\begin{multline}
    \varkappa_n\left(|\beta|\right)
    =
    \int\limits_0^{+\infty}\!\!d|\alpha|\,\,\frac{|\alpha|^{2n}}{n!} \cdot e^{-A|\alpha|^2+B|\alpha|}
    \\ 
    =  
    \frac{1}{A^{n+1/2}}\cdot\left[
    \sqrt{\frac{B^2}{4A}} 
    \cdot\!\,_1\!F_1 \left(n+1,\frac{3}{2},\frac{B^2}{4A}\right)
    \right.
    \\+\left.
    \frac{\Gamma\left(n+\frac{1}{2}\right)}{2\cdot n!} \cdot\!\,_1\!F_1\left(n+\frac{1}{2},\frac{1}{2},\frac{B^2}{4A}\right)
    \right],
    \label{kappa}
\end{multline}
%\vspace{-30pt}
where $_1\!F_1(x,y,z)$ is the Kummer confluent hypergeometric function (for large values of the third argument we utilize the approximation from Ref.\,[\onlinecite{Kumer}]), and 
\begin{gather}
    A
    =
    \frac{1}{G_1-1}+\frac{1-r_{\text{E}}}{G_2-1}+r_{\text{E}},
    \\
    B
    \equiv
    B\left(|\beta|\right)
    =
    \frac{2\gamma_a}{G_1-1}+\frac{2|\beta|\sqrt{1-r_{\text{E}}}}{G_2-1}.
\end{gather}
Function $\tilde{\varkappa}_n\left(|\beta|\right)$ is defined similarly:
\begin{multline}
    \tilde{\varkappa}_n\left(|\beta|\right)
    =
    \int\limits_0^{+\infty}\!\!d|\alpha|\,\, \frac{|\alpha|^{2n-1}}{n!}\cdot e^{-A|\alpha|^2+B|\alpha|}
    \\
    =
    \frac{1}{A^{n-1/2}}\cdot\left[
    \sqrt{\frac{B^2}{4A}}\cdot\frac{\Gamma\left(n+\frac{1}{2}\right)}{n!}\cdot\!\,
    _1\!F_1\left(n+\frac{1}{2},\frac{3}{2},\frac{B^2}{4A}\right)
    \right.
    \\+\left.
    \frac{1}{2n}\cdot\!\,_1\!F_1\left(n,\frac{1}{2},\frac{B^2}{4A}\right)
    \right].
    \label{kappa_tilde}
\end{multline}
For $n=0$ the integral from Eq.\,(\ref{kappa_tilde}) does not converge---let us address this case individually. 
Instead considering two terms of the series in Eq.\,(\ref{approx_bessel}), we take into account only the first one, which makes the primary contribution to the sum for large $|\alpha|$; thus, we get
\begin{multline}
   \varkappa_0\left(|\beta|\right)
    =
    \!\!\int\limits_{0}^{+\infty}\!\!d|\alpha|\,\,
    e^{-A|\alpha|+B|\alpha|}
    =\\
    \frac{\sqrt{\pi}e^{B^2/4A}}{2\sqrt{A}}\text{erfc}\left(\frac{B}{2\!\sqrt{A}}\right),
\quad \tilde{\varkappa}_0\left(|\beta|\right)=0. 
\label{kappa0}
\end{multline}
With the beam splitter attack, Eve gets one of two non-equiprobable quantum states: $\hat{\rho}_{\text{E}}^{(0)}$ with probability $q_0=\frac{p\left(\checkmark|0\right)}{2p_\checkmark}$ and $\hat{\rho}_{\text{E}}^{(1)}$ with $q_1=\frac{p\left(\checkmark|1\right)}{2p_\checkmark}$.
The explicit forms of diagonal $\hat{\rho}_{\text{E}}^{(0)}$ and $\hat{\rho}_{\text{E}}^{(1)}$ can be obtained by substituting the results from Eqs.\,(\ref{kappa}--\ref{kappa0}) into Eq.\,(\ref{rhodiag}).
Eve's ensemble $\mathcal{E}$ can be shortly defined as
\begin{equation}
    \mathcal{E}=\left\{\left(q_0,\,\hat{\rho}_{\text{E}}^{(0)}\right),\;\left(q_1,\,\hat{\rho}_{\text{E}}^{(1)}\right)\right\}.
\end{equation}
The maximum information that Eve can obtain about Alice's bit on average is bounded by the Holevo quantity $\chi\left(\mathcal{E}\right)$:
\begin{multline}
    I(\text{A},\text{E})\leq\chi\left(\mathcal{E}\right)
    =
    S\left(q_0\hat{\rho}_{\text{E}}^{(0)}+q_1\hat{\rho}_{\text{E}}^{(1)}\right)
    \\-
    q_0S\left(\hat{\rho}_{\text{E}}^{(0)}\right)
    -
    q_1S\left(\hat{\rho}_{\text{E}}^{(1)}\right).
    \label{eve_holevo}
\end{multline}
Since $\hat{\rho}_{\text{E}}^{(0)}$ and $\hat{\rho}_{\text{E}}^{(1)}$ are diagonal, Eq.\,(\ref{eve_holevo}) can be simplified by replacing von Neuman entropy $S$ with the classical Shannon entropy $H$
\begin{multline}
    S\left(q_0\hat{\rho}_{\text{E}}^{(0)}+q_1\hat{\rho}_{\text{E}}^{(1)}\right)
    \\=
    H\left(
    \,
    \bigg\{
    q_0\braket{n|\hat{\rho}_{\text{E}}^{(0)}|n}+q_1\braket{n|\hat{\rho}_{\text{E}}^{(1)}|n}
    \bigg\}_{n=0}^{+\infty}
    \,
    \right),
\end{multline}
\begin{equation}
    S\left(\hat{\rho}_{\text{E}}^{(a)}\right)
    =
    H\left(
    \,
    \bigg\{\braket{n|\hat{\rho}_{\text{E}}^{(a)}|n}\bigg\}_{n=0}^{+\infty}
    \,
    \right),
\end{equation}
where the Shannon entropy of a probability distribution $\left\{w_j\right\}_j$ is defined as $H\left(\{w_j\}_j\right)=-\sum\limits_{j}w_j\log_2\left(w_j\right)$.\\

\subsection*{4.4 $\quad$  Correlations}
Stealing a proportion $r_\text{E}$ of a coherent signal pulse $\ket{\gamma}$ provides Eve with a state $\ket{\sqrt{r_\text{E}}\gamma}$ uncorrelated with $\ket{\gamma}$, as the state of the joint system is described by product $\ket{\sqrt{1-r_\text{E}}\gamma}\otimes\ket{\sqrt{r_\text{E}}\gamma}$.
This is, however, not the case when the signal pulse is subject to optical amplification, turning a pure coherent state into a mixture of coherent states: now, by splitting off same $r_\text{E}$-fraction of this mixture, Eve gets a correlated state containing more information about the sent bit.
Therefore, contrary to one's executions, for Eve standing right next to Alice may be less effective than somewhere further along the line---provided that the noise from the optical amplifiers does not overweight the benefits of correlations.
Assuming that Eve performs the beam splitter attack, we quantify the correlation between Eve's and Bob's measurement results by calculating the Pearson correlation coefficient \cite{corr_coef, Pearson}
\begin{equation}
    R_{\text{BE}}
    =
    \frac{\left\langle\langle n_{\text{B}} n_{\text{E}} \right\rangle\rangle}{\sigma_{n_{\text{B}}}\cdot \sigma_{n_{\text{E}}}}
    ,
    \label{RBE}
\end{equation}
where $\langle\langle\dots\rangle\rangle$ stands for irreducible correlator, $\sigma_{n_{\text{B}(\text{E})}}$ is the standard variance.
The values are defined as follows
\begin{equation}
    \left\langle\langle n_{\text{B}} n_{\text{E}} \right\rangle\rangle
    =
    \left\langle  {n}_{\text{B}} {n}_{\text{E}}  \right\rangle_{\hat{\rho}_{\text{BE}}} - \left\langle  {n}_{\text{B}}  \right\rangle_{\hat{\rho}_{\text{B}}} \cdot \left\langle  {n}_{\text{E}}  \right\rangle_{\hat{\rho}_{\text{E}}},
\end{equation}

\begin{equation}
\begin{gathered}
    \sigma_{n_{\text{B}}}
    =
    \sqrt{
    \left\langle  {n}_{\text{B}}^2\right\rangle_{\hat{\rho}_{\text{B}}} - \left\langle  {n}_{\text{B}}  \right\rangle_{\hat{\rho}_{\text{B}}}^2
    }
    ,\\
    \sigma_{n_{\text{E}}}
    =
    \sqrt{
    \left\langle  {n}^2_{\text{E}}   \right\rangle_{\hat{\rho}_{\text{E}}} - \left\langle  {n}_{\text{E}}  \right\rangle^2_{\hat{\rho}_{\text{E}}} 
    }.
\end{gathered}
\end{equation}
Averaging over a density matrix $\hat{\rho}$ is denoted here as $\left\langle  \dots  \right\rangle_{\hat{\rho}}$.
As follows from Eq.\,(\ref{ASE_final}), the Bob-Eve density matrix can be written as  
\begin{multline}
    \hat{\rho}_{\text{BE}}
    =
    \int\!\!d^2\alpha\,\,
    P\left(\alpha,\sqrt{T_1}\gamma,G_1\right)
    \\\times\int\!\!d^2\beta\,\,
    P\left(\beta,\sqrt{T_2(1-r_{\text{E}})}\alpha,G_2\right)
    \\\times
    \int\limits_{0}^{2\pi}\!\!\frac{d\varphi}{2\pi}\,\,
    \ket{\sqrt{1-r_{\text{E}}}\alpha e^{i\varphi}}\bra{\sqrt{1-r_{\text{E}}}\alpha e^{i\varphi}}_{\text{B}}
    \\\otimes\ket{\sqrt{r_{\text{E}}}\alpha e^{i\varphi}}\bra{\sqrt{r_{\text{E}}}\alpha e^{i\varphi}}_{\text{E}},
\end{multline}
\begin{equation}
    \hat{\rho}_{\text{B}}
    =\tr_\text{E}\left[\hat{\rho}_{\text{BE}}\right],\quad
    \hat{\rho}_{\text{E}}
    =\tr_\text{B}\left[\hat{\rho}_{\text{BE}}\right],
\end{equation}
where the effective amplification coefficients $G_1$, $G_2$ and transmission probabilities $T_1$, $T_2$ are defined by Eqs.\,(\ref{G1}) and (\ref{G2}).
The average photon numbers of Bob's and Eve's subsystems are
\begin{multline}
    \!\!\!\!\!\!\left\langle {n}_\text{B}  \right\rangle _{\hat{\rho}_{\text{B}}}
    = 
    \int\!\!d^2\alpha\,\,
    P\left(\alpha,\sqrt{T_1}\gamma,G_1\right)
    \\\times\int\!\!d^2\beta\,\,
    P\left(\beta,\sqrt{T_2(1\!-\!r_{\text{E}})}\alpha,G_2\right)
    \cdot |\beta|^2
    \\=
    (1-r_{\text{E}}) \cdot \left(|\gamma|^2+G_1-1\right)+G_2-1
    ,
    \label{mean_B}
\end{multline}
\begin{multline}
    \left\langle {n}_\text{E}  \right\rangle _{\hat{\rho}_{\text{E}}}
    = 
    \int d^2 \alpha\,\,
    P \left(\alpha, \sqrt{T_1}\gamma, G_1 \right)\cdot |\sqrt{r_{\text{E}}}\alpha|^2 
    \\=
    r_{\text{E}} \cdot \left(|\gamma|^2+G_1-1 \right).
    \label{mean_E}
\end{multline}
The average product value is
\begin{multline}
    \left\langle {n}_\text{B} {n}_\text{E} \right\rangle_{\hat{\rho}_{\text{BE}}} = r_{\text{E}}(G_2-1) \cdot \left( |\gamma|^2 + G_1-1 \right)\\
    +r_{\text{E}}(1-r_{\text{E}}) \cdot \Big(2(G_1-1)^2 \\+ 4 |\gamma|^2 (G_1-1)+|\gamma|^4 \Big).
    \label{mean_BE}
\end{multline}
Hence, the expression for the irreducible correlator depends only on $|\gamma|^2, \, r_{\text{E}}$ and $G_1$:
\begin{equation}
    \left\langle\langle n_{\text{B}} n_{\text{E}} \right\rangle\rangle = r_{\text{E}}(1-r_{\text{E}}) \cdot (G_1-1)\cdot \left(2  |\gamma|^2+G_1-1 \right).
    \label{irr_cor}
\end{equation}
Here, we utilized the fact that $\langle n\rangle_{\ket{\alpha}\bra{\alpha}}=|\alpha|^2$.
In turn, the variances are obtained using the fact that $\langle n^2\rangle_{\ket{\alpha}\bra{\alpha}}=|\alpha|^4+|\alpha|^2$:
\begin{multline}
    \sigma^2_\text{B}=
    (1-r_{\text{E}})\cdot \left( |\gamma|^2+G_1-1 \right) \cdot \left( 1+2(G_2-1) \right)
    \\
    +(1-r_{\text{E}})^2 \cdot (G_1-1) \cdot \left( 2|\gamma|^2+G_1-1 \right)\\+G_2(G_2-1),
    \label{var_B}
\end{multline}
\begin{multline}
    \sigma^2_{\text{E}}=r_{\text{E}}\left(|\gamma|^2+G_1-1 \right)\\+
    r_{\text{E}}^2(G_1-1)\left(2|\gamma|^2+G_1-1 \right).
    \label{var_E}
\end{multline}
Substituting Eqs.\,(\ref{irr_cor}--\ref{var_E}) and Eqs.\,(\ref{G1}, \ref{G2}) into Eq.\,(\ref{RBE}) yields the dependence of $R_{\text{BE}}$ on $D_{\text{AE}}$.
Function $R_{\text{BE}}\left(D_{\text{AE}}\right)$ is monotonically growing which shows that with Eve approaching Bob, their measurement results become more and more correlated.
As expected, if $G_1 = 1$---corresponding to the case where Eve is right next to Alice---$R_{\text{BE}}\left(D_{\text{AE}}\right)$ vanishes, meaning zero correlations. 

\section*{NOTE 5. Phase encoding}
In this note, we study the phase encoding scheme.
We perform our analysis along the same lines as in the case of the photon number encoding.\\

\subsection*{5.1 $\quad$  Bob's information}
For self-evident reasons, phase randomization approach is inapplicable in case of phase encoding-based protocol.
The density matrix describing the tripartite system right before Bob's measurement is 
\begin{multline}
    \hat{\rho}^{\rightarrow \text{Bob}}_\text{AS} = \frac{1}{2} 
    \sum \limits_{a=0,1} \ket{a} \bra{a}_\text{A} \otimes 
    \int\!\!d^2\alpha\,\,
    P(\alpha,\sqrt{T_1} \gamma_a,G_1) 
    \\\times \left( 
    \int\!\!d^2\beta\,\,
    P \left( \beta,\sqrt{(1-r_{\text{E}})T_2} \alpha, G_2 \right) \cdot 
    \ket{\beta} \bra{\beta }_\text{S} \right).
\end{multline}
Given that Alice sent bit $a \in \{ 0,1 \}$, the probability that Bob's measurement outcome is $b \in \{0,1 \}$ can be written as
\begin{multline}
    p(b|a)
    =
    \int\frac{d^2\alpha}{\pi\left(G_1-1\right)}\,\,\cdot\exp\left(-\frac{\left|\alpha-\gamma_a\right|^2}{G_1-1}\right)
    \\\times
    \int\frac{d^2\beta}{\pi\left(G_2-1\right)}
    \exp\left(-\frac{\left|\beta-\sqrt{1-r_{\text{E}}}\alpha\right|^2}{G_2-1}\right)\\\times\braket{\beta|\hat{E}_b|\beta}.
\end{multline}
The overlap between coherent state $\ket{\beta}$ and the state with a particular value of the $\hat{q}-$quadrature is 
\begin{equation}
    \left|\braket{q|\beta}\right|^2
    =
    \sqrt{\frac{2}{\pi}}\cdot
    \exp\left(-2\left(\text{Re}[\beta]-q\right)^2\right).
\end{equation}
For successful measurement results on the Bob's side we have
\begin{equation}
\begin{gathered}
    \braket{\beta|\hat{E}_0|\beta}
    =
    \frac{1}{2}
    \int\limits_{\theta_{\text{1}}^\prime}^{\theta_{\text{2}}^\prime}
    \!\!dq\,\, e^{-2(\text{Re}[\beta]-q)^2},\\
    \braket{\beta|\hat{E}_1|\beta}
    =
    \frac{1}{2}
    \int\limits_{-\theta_{\text{2}}^\prime}^{-\theta_{\text{1}}^\prime}
    \!\!dq\,\, e^{-2(\text{Re}[\beta]-q)^2}.
\end{gathered}
\end{equation}
The expression for conditional probabilities $p(b|a)$ can be determined analytically 
\begin{multline}
    p(b|a)
    =
    \frac{1}{2}\text{erf}
    \left(
    \frac{
    \sqrt{2}\left(\theta_{\text{2}}^\prime+(-1)^{a+b}\cdot\gamma\sqrt{1-r_{\text{E}}}\right)
    }{
    \sqrt{1+2\left(G_2-1\right)+2\left(1-r_{\text{E}}\right)\left(G_1-1\right)}
    }
    \right)
    \\
    \!\!-\frac{1}{2}\text{erf}
    \left(
    \frac{
    \sqrt{2}\left(\theta_{\text{1}}^\prime+(-1)^{a+b}\cdot\gamma\sqrt{1-r_{\text{E}}}\right)
    }{
    \sqrt{1\!+\!2\left(G_2\!-\!1\right)+2\left(1\!-\!r_{\text{E}}\right)\left(G_1\!-\!1\right)}
    }
    \right).
\end{multline}
Since $p(\checkmark|0)=p(\checkmark|1)$, the mutual information between Alice and Bob is 
\begin{equation}
    I\left(\text{A},\text{B}\right)
    =
    S(\text{A})-S(\text{A}|\text{B})
    =
    1-h_2\left(\frac{p(1|0)}{p_\checkmark}\right).
\end{equation}
As a result, we obtain the explicit form of the expression $S(\text{A})-S(\text{A}|\text{B})$ which we substitute into Eg.\,(6) of the main text.

\subsection*{5.2 $\quad$  Eve's information}
The density matrix of the Alice-Bob-Eve system given that Bob obtained conclusive measurement result is
\begin{multline}
    \hat{\rho}_{\text{ABE}}^{\text{f}}
    =
    \frac{1}{2}
    \sum \limits_{a=0,1} 
    \frac{1}{p(\checkmark|a)}\ket{a} \bra{a}_\text{A}
    \otimes
    \sum\limits_{b=0,1}
    \ket{b}\bra{b}_{\text{B}}
    \\\otimes
    \int\!\!d^2\alpha\,\, P(\alpha,\sqrt{T_1}\gamma_a,G_1)
    \cdot
    \int\!\!d^2\beta\,\,P\left( \beta,\sqrt{(1\!-\!r_{\text{E}})T_2} \alpha, G_2 \right)
    \\\times\braket{\beta|\hat{E}_b|\beta}
    \ket{\sqrt{r_{\text{E}}}\alpha}\bra{\sqrt{r_{\text{E}}}\alpha}_\text{E}.
\end{multline}
To estimate the conditional entropy $S(\text{A}|\text{E})$ one has to calculate the reduced Alice-Eve density matrix by tracing out Bob's subsystem:
\begin{multline}
    \hat{\rho}_{\text{AE}}^{\text{f}}
    =
    \text{tr}_{\text{B}}\left[
    \hat{\rho}_{\text{ABE}}^{\text{f}}
    \right]
    =
    \frac{1}{2}\sum\limits_{a=0,1}
    \frac{1}{p(\checkmark|a)}\ket{a}\bra{a}_\text{A}
    \\\otimes
    \int\!\!d^2\alpha\,\,
    P\left(\alpha,\sqrt{T_1}\gamma_a,G_1\right)
    \cdot
    \\\times\int\!\!d^2\beta\,\,P\left( \beta,\sqrt{(1-r_{\text{E}})T_2} \alpha, G_2 \right)
    \\\times\braket{\beta|\hat{E}_\checkmark|\beta}
    \ket{\sqrt{r_{\text{E}}}\alpha}\bra{\sqrt{r_{\text{E}}}\alpha}_\text{E}.
\end{multline}
It can be rewritten as
\begin{equation}
    \hat{\rho}_{\text{AE}}^{\text{f}}
    =
    \int\!\!d^2\alpha\,\,
    Q_{\checkmark}[\alpha]\cdot\hat{\rho}_{\text{AE}}^{\text{f}}\left[\alpha\right],
\end{equation}
where
\begin{multline}
\!\!\!\!
    Q_{\checkmark}[\alpha]\!
    =\!
    \frac{P\left(\alpha; \!\sqrt{T_1}\gamma,G_1\right)}{p_\checkmark}\!\!\!
    \\\times\int\!\!d^2\beta\,\,
    P\left(\beta;\alpha\sqrt{(1\!-\!r_{\text{E}})T_2},G_2\right)\braket{\beta|\hat{E}_{\checkmark}|\beta},\!\!\!
\end{multline}
\begin{multline}
    \hat{\rho}_{\text{AE}}^{\text{f}}\left[\alpha\right]
    =
    \frac{1}{2}\sum\limits_{a=0,1}
    \ket{a}\bra{a}_\text{A}\\\otimes\left|(-1)^a\sqrt{r_{\text{E}}}\alpha\right\rangle\left\langle(-1)^a\sqrt{r_{\text{E}}}\alpha\right|_\text{E}.
    \label{alpha_density}
\end{multline}
Mutual information between the eavesdropper and Alice after the post-selection procedure is
\begin{equation}
    I\left(\text{A},\text{E}\right)
    =
    S(\text{A})-S(\text{A}|\text{E})
    =
    1-S(\text{A}|\text{E}).
\end{equation}
The lower bound on the entropy $S(\text{A}|\text{E})$ may be found by exploiting the concavity of conditional quantum entropy %\cite{?}
\begin{multline}
    S(\text{A}|\text{E})
    \geq 
    \int \!\! d^2\alpha\,\,
    Q_{\checkmark}[\alpha]\cdot
    S_{\hat{\rho}_{\text{AE}}^{\text{f}}[\alpha]}(\text{A}|\text{E})
    \\=
    \int\!\!d^2\alpha\,\, Q_{\checkmark}[\alpha] \cdot \left[1-h_2\left( \frac{1+\left|\braket{-\sqrt{r_{\text{E}}}\alpha|\sqrt{r_{\text{E}}}\alpha}\right|}{2} \right)\right]
    \\=
    1-\int\!\!d^2\alpha\,\, Q_{\checkmark}[\alpha] \cdot h_2\left( \frac{1+\exp\left(-2r_{\text{E}}|\alpha|^2\right)}{2} \right),
\end{multline}
where $S_{\hat{\rho}_{\text{AE}}^{\text{f}}[\alpha]}(\text{A}|\text{E})$ denotes conditional entropy for Alice-Eve density matrix described by Eq.\,(\ref{alpha_density}).
Utilizing Jensen's inequality,
\begin{equation}
    \left\langle h_2(x)\right\rangle
    \leq
    h_2\left(\left\langle x\right\rangle\right),
\end{equation}
we bound Eve's information as
\begin{equation}
    I(\text{A},\text{E})
    \leq h_2\left(\frac{1+\left\langle\exp\left(-2r_{\text{E}}|\alpha|^2\right)\right\rangle_{Q_{\checkmark}}}{2}\right).
    \label{eve_inf}
\end{equation}
Straightforward calculations allows us to find
\begin{multline}
    \left\langle \exp\left(-2r_{\text{E}}|\alpha|^2\right) \right\rangle_{Q_{\checkmark}}
    = 
    \int\!\!d^2\alpha\,\,
    Q_{\checkmark}[\alpha]\cdot e^{-2r_{\text{E}}|\alpha|^2}
    \\= 
    \left[ \sum\limits_{x=0,1}\text{erf} \left( \frac{\theta_{\text{2}}^\prime \cdot \left[1 \! + \! 2r_{\text{E}}(G_1 \! - \! 1) \right]\! +(-1)^x \! \sqrt{(1\! - \! r_{\text{E}})}\gamma}{ \zeta \cdot \sqrt{1\!+ \! 2r_{\text{E}}(G_1\!- \!1)}} \right) \right.
    \\- \! \left.
    \sum\limits_{x=0,1}\text{erf} \left( \frac{\theta_{\text{1}}^\prime \cdot \left[1 \! + \! 2r_{\text{E}}(G_1 \! - \! 1) \right]\! +(-1)^x \! \sqrt{(1\! - \! r_{\text{E}})}\gamma}{ \zeta \cdot \sqrt{1\!+ \! 2r_{\text{E}}(G_1\!- \!1)}} \right)
    \right]
    \\\times
    \frac{\exp\left(\frac{-2r_{\text{E}}|\gamma|^2}{1+2r_{\text{E}}(G_1-1)}\right) }{ 2p_\checkmark \left(1 \! + \! 2r_{\text{E}}(G_1 \! - \! 1)\right) },
    \label{av_exp}
\end{multline}
where $\zeta=\sqrt{G_1+G_2+2r_{\text{E}}(G_1-1)(G_2-1)-3/2}$.
By substituting Eq.\,(\ref{av_exp}) into Eq.\,(\ref{eve_inf}) we obtain the upper bound for Eve’s information ($I(\text{A:E})$ in the Eq.\,(6) of the main text).

%%%%%%%%%%%%%%%%%%%%%%%%%%%%%%%%%%%%%%%%%%%%%%%%%
%%%%%%%%%%%% BIBLIOGRAPHY (SI) %%%%%%%%%%%%%%%%%%
%%%%%%%%%%%%%%%%%%%%%%%%%%%%%%%%%%%%%%%%%%%%%%%%%

\end{document}